\documentclass[a4paper,11pt]{article}
\pdfoutput=1 

\usepackage{jcappub} 

\usepackage{multirow}

\title{Cosmic-Ray Positrons Strongly Constrain Leptophilic Dark Matter}
\author{Isabelle John}
\author{and Tim Linden}
\affiliation{Stockholm University and The Oskar Klein Centre for Cosmoparticle Physics, Alba Nova, 10691 Stockholm, Sweden}
\emailAdd{isabelle.john@fysik.su.se, ORCID: orcid.org/0000-0003-2550-7038}
\emailAdd{linden@fysik.su.se, ORCID: orcid.org/0000-0001-9888-0971}

\abstract{Cosmic-ray positrons have long been considered a powerful probe of dark matter annihilation. In particular, myriad studies of the unexpected rise in the positron fraction have debated its dark matter or pulsar origins. In this paper, we instead examine the potential for extremely precise positron measurements by AMS-02 to probe hard leptophilic dark matter candidates that \emph{do not} have spectral features similar to the bulk of the observed positron excess. Utilizing a detailed cosmic-ray propagation model that includes a primary positron flux generated by Galactic pulsars in addition to a secondary component constrained by He and proton measurements, we produce a robust fit to the local positron flux and spectrum. We find no evidence for a spectral bump correlated with leptophilic dark matter, and set strong constraints on the dark matter annihilation cross-section that fall below the thermal annihilation cross-section for dark matter masses below 60~GeV and 380~GeV for annihilation into $\tau^+\tau^-$ and $e^+e^-$, respectively, in our default model.}

\begin{document}
\maketitle
\flushbottom

\section{Introduction}
The AMS-02 experiment, onboard the International Space Station, has provided extremely precise observations of the flux, composition and spectrum of local cosmic-rays~\cite{AMSexperiment}. These data yield new insights into the energetics of the Galaxy's most powerful sources, and the properties of Galactic transport. Additionally, searches for unexpected spectral features can potentially provide evidence for new physics, such as anisotropic diffusion or even dark matter annihilation.

Notably, AMS-02 observations have built upon data by the PAMELA observatory \cite{Adriani:2010rc} and earlier instruments \cite{Barwick:1997ig, 1987A&A...188..145G}, to provide strong evidence for a steep rise in the positron fraction above 10~GeV, and a drop-off at 300~GeV~\cite{PhysRevLett.122.041102}. The two most plausible explanations for this excess are pulsars, e.g.,~\cite{Hooper:2008kg, Profumo:2008ms, Linden:2013mqa, Yuksel:2008rf, Malyshev:2009tw}, and annihilating dark matter (DM) particles, e.g.,~\cite{Turner:1989kg, Arkani-Hamed:2008hhe, Cholis:2013psa, Bergstrom:2008gr, Cirelli:2008jk, Cholis:2008hb, Cirelli:2008pk, Cholis:2008qq, Cholis:2008wq}. Recently, observations of bright \mbox{$\gamma$-ray} emission surrounding young and middle aged pulsars (e.g., Geminga, Monogem, and a handful of younger HAWC and H.E.S.S. sources that are now named TeV halos~\cite{Abeysekara:2017hyn, Linden:2017vvb, HAWC:2017kbo, HESS:2018pbp, DiMauro:2019hwn, Sudoh:2019lav, Johannesson:2019jlk, Sudoh:2021avj}), which are powered by the efficient acceleration of high-energy electrons and positrons, have favored the pulsar interpretation~\cite{Hooper:2008kg, Yuksel:2008rf, Profumo:2008ms, Malyshev:2009tw, Grasso_2009, Linden:2013mqa, Hooper:2017gtd, Hooper:2017tkg, Fang:2018qco, Profumo:2018fmz, Manconi:2020ipm}. Hints for a pulsar contribution have also been found in the cosmic-ray electron spectrum \cite{DiMauro:2020cbn}. At the same time, dark matter models have come into tension with $\gamma$-ray limits from several targets, including dwarf spheroidal galaxies \cite{Ackermann_2011, Cholis:2013psa, Geringer_Sameth_2011, Cholis_2012, Hoof_2020, Albert_2017, Ackermann_2015}, although recent studies of systematic issues in both J-factor estimations and $\gamma$-ray analysis have found several complications that could significantly relax this tension \cite{Calore:2018sdx, Alvarez:2020cmw, Ando:2020yyk}. Other explanations consider that the positron excess consists entirely of secondary positrons powered by supernova activity \cite{Mertsch:2009ph, Blum:2013zsa, Kohri:2015mga}.

\begin{figure}[tbp]
\centering
\includegraphics[width=1\textwidth]{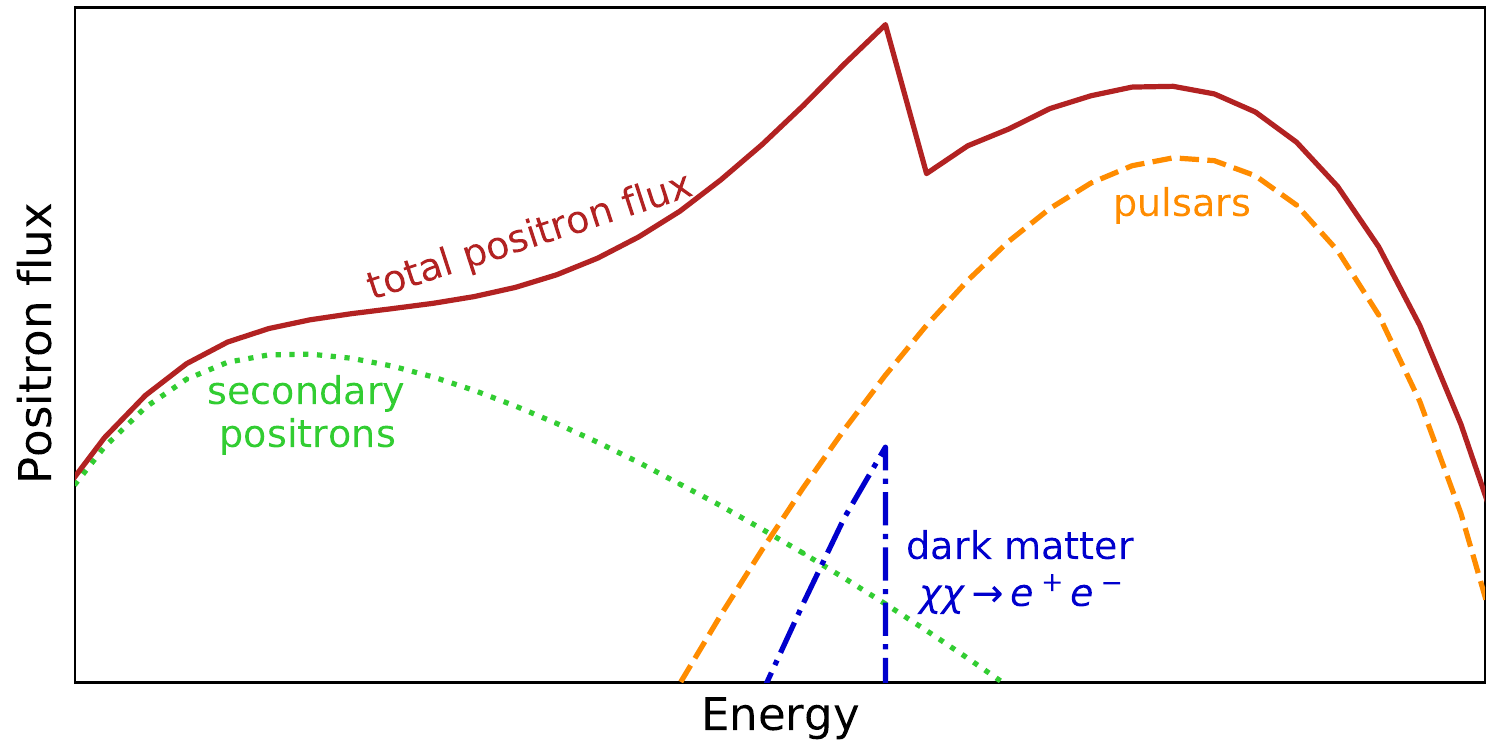}
\caption{A schematic diagram of the scenario we examine. We model a positron flux that includes a primary contribution from pulsars, secondary contributions from the hadronic interactions of primary cosmic-rays, and a highly peaked contribution from DM annihilation into leptonic final states, which we attempt to constrain. DM signals detectable by AMS-02 generally range from a few GeV to a few hundred GeV.}
\label{fig: DM_example}
\end{figure}

In this paper, we consider the constraints that can be placed on DM in scenarios that attribute the bulk of the rising positron fraction above 10~GeV to Galactic pulsars. We make use of the high-statistical precision of the AMS-02 data to calculate strong constraints on DM annihilation into highly-peaked leptonic final states, which have spectra that are inconsistent with the high-energy positron spectrum. A schematic example of the scenario we examine is shown in Figure~\ref{fig: DM_example}: At lower energies, the positron flux is dominated by secondary positrons produced by the interaction of protons and He nuclei with the interstellar medium, while at higher energies, most positrons are produced in electron-positron pairs accelerated by pulsars. Against this background, our analysis probes a subdominant contribution from annihilating DM.

DM annihilation into hard-spectrum leptonic final states are well-motivated in the context of leptophilic dark matter models~\cite{Arkani-Hamed:2008hhe, Bell:2014tta}. Originally discussed in the context of fitting cosmic-ray excesses from PAMELA, AMS-02, ATIC and DAMA/LIBRA~\cite{Fox:2008kb, Kopp:2009et}, leptophilic models have also been theoretically motivated, for example, in the context of Z' models~\cite{Bell:2014tta}, and have been found to produce unique signatures in particle colliders~\cite{Freitas:2014jla, delAguila:2014soa, Chen:2015tia}. Moreover, models with significant leptonic couplings are generically expected in well-motivated scenarios such as dark photon models~\cite{Pospelov:2007mp, Alexander:2016aln}. We also note that signals from hard-spectrum leptonic final states may also be dominant even in models with loop-level hadronic interactions~\cite{DEramo:2017zqw}.

Our result builds on a similar analysis by Bergstr\"om et al.~(2013) \cite{Bergstrom:2013jra}. Their work tests two discrete models for the astrophysical background, as well as contributions from DM particles between 5--300~GeV annihilating into leptonic final states ($\tau^+\tau^-$, $\mu^+\mu^-$, $e^+e^-\gamma$ and $e^+e^-$). Their strongest limits are obtained for the direct $e^+e^-$ channel, where they set upper limits on the DM velocity-averaged annihilation cross section of about $10^{-28}$~cm$^3$~s$^{-1}$ at 5~GeV and about $10^{-25}$ cm$^3$~s$^{-1}$ at 250~GeV.

In this paper, we update and improve on the Bergstr\"om et al. results in three ways:
(1) Our analysis utilizes a larger dataset provided by the latest AMS-02 measurements of positron, proton, and He fluxes \cite{PhysRevLett.122.041102, PhysRevLett.121.051101, PhysRevLett.114.171103, PhysRevLett.115.211101}.
(2) We use the cosmic-ray propagation code \texttt{Galprop} \cite{Galprop} to create a full cosmic-ray propagation model for the positron, proton and He data. Our fit allows a large variety of parameters, such as the diffusion and particle injection spectra in the fit to float, while Bergstr\"om et al. build their background from a handful of pre-defined \texttt{Galprop} models. (3) We implement a new solar modulation model \cite{Cholis:2020tpi} to describe the effect of solar activity on low-energy cosmic-ray particles, rather than approximating solar modulation by a force-field potential.

The paper is outlined as follows: In Section~\ref{sec:models}, we describe the methodology of this analysis and explain the data selection~(Sec.~\ref{subsec:dataselection}), the computational model~(Sec.~\ref{subsec:computational model}), the DM model~(Sec.~\ref{subsec:calculation of DM limits}) and the statistical analysis (Sec.~\ref{subsec:statistical analysis}). The results are presented in Section~\ref{sec:results}, and our conclusions are discussed in Section~\ref{sec:discussion}.

\section{Methodology}
In this section, we describe our selection of the AMS-02 data, and present our methodology for constructing astrophysical models that fit this data. We also describe our statistical analysis to constrain the DM contribution.

Our analysis generally proceeds as follows. To compute the DM limits, we perform a fit of the positron, proton and He data from AMS-02 to constrain the parameters of our astrophysical model. While the high-energy positron spectrum is dominated by primary positrons that are created by pulsars as electron-positron pairs, low-energy positrons are produced as secondaries from proton interactions with a subdominant contribution from He (about 10\%) \cite{PhysRevLett.115.211101}. After obtaining an astrophysical model for the positron flux in the absence of DM, we add a DM contribution and compute the 95\% confidence upper limit on the DM annihilation rate. Notably, we allow critical astrophysical parameters to be re-fit for each DM mass and cross-section, producing a more realistic constraint on the DM contribution. In the following sections, the procedure is described in more detail.

\label{sec:models}
\subsection{Data Selection}
\label{subsec:dataselection}
Our analysis takes advantage of the recently published AMS-02 data for positrons \cite{PhysRevLett.122.041102}, protons \cite{PhysRevLett.121.051101, PhysRevLett.114.171103}, and He \cite{PhysRevLett.115.211101}. A summary of the data sets is given in Table~\ref{tab: AMSdata}. We note that our constraints on leptophilic DM only directly depend on our analysis of the positron data, but we include protons and He in the astrophysical model to obtain better constraints on the model parameters. We also note that protons and He have the same electric charge as positrons, which decreases the relative uncertainties from solar modulation.

For positrons, we consider energies in the range from 2 to 1000~GeV in our fit. We do not include energies below 2~GeV due to their large uncertainties, especially coming from solar modulation \cite{Gleeson:1968zza}, and because the DM contribution is small at low energies compared to the very bright contribution from secondary production (see e.g. Figure~\ref{fig: DM_example} for the different contributions to the positron flux). To achieve the best statistical precision, we take a large dataset that includes the AMS-02 data from May 2011 to November 2017 \cite{PhysRevLett.122.041102}. For protons, we consider energies in the range from 2 to 1800~GeV, for which we combine two data sets: From 2 to 60~GeV, AMS-02 provides time-dependent data \cite{PhysRevLett.121.051101}, and we select the data from February 2015 to May 2017 (Bartels rotations 2477--2506), after the polarity flip of the solar magnetic field was completed, to avoid additional uncertainties from the effects of the solar modulation. At energies from 60 to 1800~GeV, we use the time-independent data in \cite{PhysRevLett.114.171103}, recorded from May 2011 to November 2013. At these energies, the effect of solar modulation is negligible. For He, we consider the energy range from 2 GeV to 1000~GeV and use the time-independent data from May 2011 to November 2013, which includes both $^3$He and $^4$He \cite{PhysRevLett.115.211101}. As the conversion from rigidity to energy depends on the number of nucleons, we estimate the ratio of $^3$He to $^4$He in this flux by linearly extrapolating the ratio given in~\cite{PhysRevLett.123.181102} to fluxes at higher energies.

\begin{table}[tbp]
\centering
\begin{tabular}{|c|c|c|c|} 
\hline
Particle & Energy range [GeV] & Time range & ~~Ref.~~ \\
\hline\hline
Positrons ~~& 2 -- 1000 & ~~May 2011 -- Nov. 2017~~ & \cite{PhysRevLett.122.041102} \\
\hline
\multirow{2}{*}{Protons} & 2 -- 60 & Feb. 2015 -- May 2017 & \cite{PhysRevLett.121.051101} \\
    & 60 -- 1800 & May 2011 -- Nov. 2013 & \cite{PhysRevLett.114.171103}\\
\hline
He  & 2 -- 1000 & May 2011 -- Nov. 2013 & \cite{PhysRevLett.115.211101} \\
\hline
\end{tabular}
\caption{AMS-02 data sets used in this analysis. Since our solar modulation model is time-dependent, we are able to utilize datasets from different time periods.}
\label{tab: AMSdata}
\end{table}

\subsection{Computational Model}
\label{subsec:computational model}
To simulate galactic cosmic-ray propagation, we use \texttt{Galprop~ v.56~}\cite{Galprop, Strong:1998pw, Strong:1999sv, Strong:2001fu, Strong:2009xj, Moskalenko:1997gh, Moskalenko:1999sx, Orlando:2017tde} \footnote{Note that Bergstr\"om et. al use v54. However, we do not expect this to cause any significant disparities between the results.}, a state-of-the-art tool to compute cosmic-ray propagation, that takes into account numerous processes, such as diffusion, spallation, energy losses and reacceleration. For our analysis, it is sufficient to solve the propagation equation in 2D, and to include only cosmic-ray contributions up to $^4$He, while ignoring heavier nuclei.

Galprop simulations begin by injecting a significant flux of primary cosmic-rays (most importantly for our analysis, protons and $^4$He) following the standard supernova spatial distribution~\cite{Johannesson:2015qqi}. The subsequent propagation and hadronic interactions of these primary particles produce a secondary flux, which includes more protons as well as new components (most importantly, positrons and $^3$He). We supplement these contributions from supernovae with an additional component of primary electrons and positrons powered by pulsar activity, as described below.

We tune the following Galprop parameters in our model (for a list, see Table~\ref{tab: parameters}). We model the diffusion spectrum with a diffusion coefficient $D_0$ set at a reference rigidity of 4~GV and a diffusion break $D_\text{break}$, where $\delta_1$ and $\delta_2$ are the spectral indices below and above the break, respectively. The proton injection spectrum is modeled by three free parameters, including two spectral indices and a critical energy describing their transition. This break primarily affects the lowest energy protons and positrons. Similarly, for He we introduce a spectral break and spectral indices $\gamma_1^\text{He}$ and $\gamma_2^\text{He}$ below and above the break, respectively. At higher energies, most of the positron contribution comes from pulsars, where the free parameters are the spectral index $\gamma_\text{psr}$, the cutoff energy $E_\text{cut}^\text{psr}$ of the spectrum and the pulsar formation rate $\dot{N}_{100}$ (discussed below). Reacceleration is taken into account by the Alfv\'en velocity which is most important for positrons and has a smaller effect on protons. The change in convection velocity perpendicular to the Galactic plane $dv/dz$ (with convection velocity $v=0$ at $z=0$) and the solar modulation parameters ($\phi_0$ and $\phi_1$, discussed below) are significant at lower energies. Lastly, there is an overall normalization of the spectra computed by Galprop which is related to uncertainties in the Milky Way supernova rate \cite{Diehl:2006cf}. Since secondary positrons are produced mostly by protons, the normalization is the same for both fluxes.

A particularly important parameter in the model is the halo height, $z$. The DM contribution is proportional to the halo height, and therefore we examine this parameter individually. We compute the astrophysical model for two fixed values of the halo height: setting $z=5.6$~kpc as a standard height (default model), and $z=3$~kpc as the lower limit on the halo height (small-height model) \cite{Korsmeier:2021brc, Weinrich:2020ftb, Evoli:2019iih}.

In our model, we introduce pulsars as a new source of primary positrons by modifying the Galprop code. For this, we adopt the standard pulsar distribution profile from~\cite{Lorimer:2003qc, Lorimer:2006qs}, to obtain the 3-dimensional pulsar distribution in the Galaxy,

\begin{equation}
\label{eq: pulsar profile}
\rho(R, z) = 64.6 \text{ kpc}^{-4.35} R^{2.35} \exp\left[\frac{-R}{R_0}\right] \exp\left[\frac{-|z|}{z_0}\right],
\end{equation}
where $z$ is the halo height of the Milky Way, and $R_0=1.528$~kpc and $z_0=0.330$~kpc. We model the $e^+e^-$ spectrum from pulsars as a power-law with an exponential cutoff~\cite{Hooper:2008kg}, 

\begin{equation}
\label{eq: pulsar spectrum}
\frac{dN}{dE} \approx C \left(\frac{E}{\text{GeV}}\right)^{-\gamma_{\text{psr}}} \exp\left[\frac{-E}{E_{\text{cut}}^\text{psr}}\right] \left(\text{GeVs}\right)^{-1},
\end{equation}
where $\gamma_{\text{psr}}$ is the spectral index and $E_{\text{cut}}^\text{psr}$ the cutoff energy. For the normalization, we calculate $C$ such that the average total pulsar spindown luminosity is 10$^{49}$~erg~\cite{Hooper:2008kg}, and the average efficiency for the conversion of pulsar spindown power into $e^+e^-$ pairs is 1\%, a value which is consistent with previous empirical fits to the positron data~\cite{Cholis:2018izy} and forms an intermediate value between the $\sim$10\% injection efficiency of mature TeV halos such as Geminga~\cite{Hooper:2017gtd} and the $<$1\% efficiency of young pulsars such as Vela~\cite{Sudoh:2019lav}. In our model, we let $\gamma_\text{psr}$, $E_{\text{cut}}^\text{psr}$ and $\dot{N}_{100}$ free, fitting the spectrum and normalization of the pulsar contribution.

Cosmic-ray measurements are affected by the heliosphere of the Sun, causing particle losses at low energies ($\lesssim$~10~GeV). This effect strongly depends on the orientation of the solar magnetic field and solar activity. Often (e.g., as in \cite{Bergstrom:2013jra}), the solar modulation is approximated by a force-field potential \cite{Gleeson:1968zza}. Here, we implement the solar modulation model from \cite{Cholis:2020tpi} that takes into account the time-, charge- and rigidity dependence of the solar modulation. In the solar modulation model, we choose time periods that match the data analysis periods from AMS-02, and keep the parameters $\phi_0$ and $\phi_1$, that describe the heliospheric potential, as free parameters.

Additionally, there are several important parameters that we fix to their standard Galprop values. Most notably, we utilize a Galactic magnetic field model based on Model C of Ref.~\cite{Fermi-LAT:2014ryh}, utilizing the re-fit value for the radial scale height of the regular magnetic field of 13~kpc. This model yields a magnetic field at the solar position of 5.7~$\mu$G. All parameters for our astrophysical models can be found in Table~\ref{tab: parameters} for the default model and small-height model.

\begin{table}[tbp]
\centering
\begin{tabular}{|c|c|c|c|} 
\hline
Parameter  & Range & Default & Small-height\\  
\hline\hline
\rule{0pt}{0.35cm}
Diffusion coefficient, $D_0$ [cm$^2$/s]                      & $10^{28} - 10^{30}$              & 1.636 $\cdot10^{28}$  & $1.679\cdot10^{28}$ \\
Diff. spectrum break, $D_\text{break}$ [MV]~             & $10^3 - 1.4 \cdot 10^4$          & 6.067 $\cdot 10^3$    & $6.965\cdot 10^3$       \\
Spectral index below break, $\delta_1$                       & $0.01 - 0.61$                    & 0.0527                & 0.155                   \\
Spectral index above break, $\delta_2$                       & $0.01 - 0.61$                    & 0.361                 & 0.403                   \\
Change in $v_\text{conv}$, $dv/dz$ [km/(s kpc)]                      & $0 - 10$                         & 6.345                 & 6.656                   \\
Alfv\'en velocity, $v_\text{Alfv\'en}$ [km/s]                & $1 - 30$                         & 4.524                 & 8.751                   \\
$p$ injection spectrum break [MV]                         & ~~$5\cdot 10^2 - 1\cdot 10^5$~~  & 5.195$\cdot10^2$      & $2.986\cdot10^3$        \\
$p$ spectral index below break, $\gamma_1^p$              & $1.5 - 3.0$                      & 1.657                 & 1.999                   \\
$p$ spectral index above break, $\gamma_2^p$              & $1.5 - 4.5$                      & 2.523                 & 2.459                   \\
Pulsar spectral index, $\gamma_\text{psr}$                   & $0 - 2.5$                        & 1.337                 & 1.394                   \\
Pulsar cutoff energy, $E_\text{cut}^\text{psr}$ [GeV]        & $5\cdot 10^2 - 2\cdot 10^3$      & 535.587               & 559.470                 \\
Pulsar form. rate, $\dot{N}_{100}$ [psr/century]        & $0.161 - 48.61$                  & 0.297                & 0.386                   \\
Solar mod. parameter, $\phi_0$ [GV]                    & $0.21 - 0.43$                    & 0.378                 & 0.370                   \\
Solar mod. parameter, $\phi_1$ [GV]                    & $1.15 - 1.95$                    & 1.950                 & 1.950                   \\
Normalization (positrons, protons)                           & $0.5 - 1.5$                      & 0.814                 & 0.820                   \\
\hline
\rule{0pt}{0.35cm}
He injection spectrum break [MV]                         & $\ast$      & $3.053\cdot10^5$      & $1.361\cdot10^4$        \\
He spectral index below break, $\gamma_1^\text{He}$      & $1-4$                            & 2.505                 & 2.311                   \\
He spectral index above break, $\gamma_2^\text{He}$      & $1-4$                            & 2.425                 & 2.423                   \\
Normalization (He)                                       & $ 0.5-1.5$                       & 1.101                 & 1.125                   \\
\hline
\end{tabular}
\caption{List of free parameters and the parameter ranges, as well as the best-fit values for the fit with positrons, protons and He with iMinuit (Step 2) for the default model with a 5.6~kpc halo height and the small-height model with a 3.0~kpc halo height. These models provide a good fit to the data, with a total $\chi^2$ of 88.6 over 141 d.o.f (default model) and 97.08 over 141 d.o.f. (small-height model). Comparing the small-height model to the default model, most parameters remain relatively consistent. However, we required a lower-spectral break for the He injection spectrum than allowed in the range of our default model (default model: $5\cdot 10^4 - 5\cdot 10^5$ MV, small-height model: $5\cdot 10^3 - 5\cdot 10^5$ MV). This model still provides a good fit to the data ($\chi^2$/d.o.f. is less than 1). However, the small-height model provides a worse fit ($\Delta\chi^2$~=~8.5) compared to our default model, and thus represents the minimum halo height that appears consistent with AMS-02 data.}
\label{tab: parameters}
\end{table}

\begin{table}[tbp]
\centering
\begin{tabular}{|c|c|} 
\hline
Parameter & Value \\
\hline\hline
\rule{0pt}{0.3cm}
Local DM density, $\rho_\text{loc}$ [GeV/cm$^3$] & 0.4 \\
DM halo core radius [kpc] & 20 \\
Magnetic field at Earth [$\mu$G]&  5.7\\
\hline
\end{tabular}
\caption{List of parameters that are fixed in the model.}
\label{tab: fixed parameters}
\end{table}

\subsection{Dark Matter Model}\label{subsec:calculation of DM limits}
For the DM model, we assume a local DM density of 0.4~GeV/cm$^3$ and a core radius of the DM halo of 20~kpc~(c.f. \cite{Read:2014qva}), see Table~\ref{tab: fixed parameters}, and describe the DM distribution in the Galaxy using a generalized NFW profile~\cite{Navarro:1995iw}. We utilize a non-standard NFW slope of 0.5, but note that this choice has almost no effect on the local DM annihilation rate -- producing limits that are more conservative by a few percent. We consider four different leptophilic final states for the annihilation of Majorana DM: $\chi\chi\rightarrow\tau^+\tau^-$, $\chi\chi\rightarrow\mu^+\mu^-$, $\chi\chi\rightarrow e^+e^-$, and $\chi\chi\rightarrow\phi\phi\rightarrow e^+e^-e^+e^-$, where $\phi$ is a light mediator. For the $\tau^+\tau^-$, and $\mu^+\mu^-$ final states, we use DM spectra obtained with DarkSUSY~ v.5.1.1~\cite{Gondolo:2004sc}. For the $e^+e^-e^+e^-$ and $e^+e^-$ states, we utilize analytic calculations of the DM spectrum that we directly implement in Galprop. Specifically, in the $e^+e^-e^+e^-$ case, the contribution to the differential positron flux is constant and integrates to the DM mass over the full energy range, following a similar calculation in \cite{Fortin:2009rq}. The $e^+e^-$ final state is represented by a delta function and we smear out the contribution to the positron flux over the neighboring energy bins in Galprop. Here, we neglect electroweak corrections as these are in the order of 1\%.

\subsection{Statistical Analysis}\label{subsec:statistical analysis}
To calculate the best-fit values for our model parameters, we use two minimization codes: \texttt{pyMultinest v.2.10} \cite{pymultinest, Feroz:2007kg, Feroz:2008xx, Feroz:2013hea} and \texttt{iMinuit v.1.4.3.} \cite{iminuit_website, iminuit}.

We create our astrophysical model in two steps: First we fit positrons and protons over a wide range of parameters. For this we use Multinest as it can handle the large number of free parameters well and is reliable at finding the global minimum. Then we add He to the fit and re-fit the previous parameters together with the He injection spectrum. For this, we use iMinuit as it finds the exact values of the minimum more precisely than Multinest. Because their contribution to the positron flux is negligible, we do not include nuclei heavier than He in Galprop.

Throughout our analysis, we compare our model with the AMS-02 particle fluxes in multiple energy bins. We note that the AMS-02 data includes uncertainties that stem from both statistical and systematic errors. The systematic errors are likely to be correlated between energy bins, but to date, the AMS-02 collaboration has not released a covariance matrix that accounts for the correlation between systematic errors in nearby energy bins. Thus, in this paper, we treat the systematic and statistical errors in each bin independently, adding them in quadrature. This leads to an overestimate of the \mbox{AMS-02} error budget, and allows our models to achieve goodness of fit values less than 1. Notably, however, this preserves the same minimum value for the best-fit, and still allows us to compare models in terms of their $\Delta\chi^2$.

For the first step, we use Multinest to fit the astrophysical parameters corresponding to proton and positron propagation discussed in Section~\ref{subsec:computational model}. We show the intermediate results of this procedure in Appendix~\ref{sec:multinest}. 

For the second step, we take the best-fit values and perform an additional fit by adding the He data to the model, using iMinuit. We set the $^4$He injection spectrum to follow a twice-broken power-law, with a fixed spectral index below 5~GV of 1.8, and allow the two higher spectral indices and the location of the higher spectral break to float. We allow the normalization of the primary He flux to float. We note that there is also a small primary $^3$He flux, with a spectrum set to the primary proton spectrum. The relative normalization of these terms, $7.199\cdot10^4$ to 9.033, makes the effect of primary $^3$He injection negligible. Note that in Galprop, these abundances are defined relative to the (arbitrarily chosen) proton normalization of $1.06\cdot10^6$. We perform the fit to the He data using iMinuit, also allowing the parameters of the previous step to float. As we want to focus on the positron fit of our model, we introduce an additional systematic uncertainty of $2\%$ into the He data, accepting that our model does not include several uncertainties that are specific to He. More importantly, the systematic uncertainties in the He flux are correlated, but here we treat them as uncorrelated.

For the third step, we investigate the DM contribution. While we include protons and He when fitting our astrophysical background model to the data, to obtain more realistic constraints on the positron flux, we consider the combined (DM and astrophysical contribution) to only the positron flux when we set DM limits. This always produces more conservative upper limits because DM does not directly contribute to the proton or He fluxes (leaving the secondary positron contribution approximately the same), and thus the combined (proton+He+positron) $\chi^2$ fit typically provides stronger limits if DM induced adjustments to the diffusion parameters perturb these datasets. 

To determine the DM limits, we study DM masses between 5 and 2000~GeV for different annihilation cross sections $\langle\sigma v\rangle$. For each DM mass and final state, we set up a grid of annihilation cross sections using logarithmic steps in the mass range of interest and an iterative process to obtain values for the annihilation cross section, and fit the model using five parameters that are particularly relevant for the DM contribution to the positron flux ($D_0$, $\delta_2$) or are degenerate with the DM contribution ($E^\text{psr}_\text{cut}$, $\gamma_\text{psr}$ and $\dot{N}_{100}$), and let them float as well for each grid point. Then we compute the $\chi^2$ profile of $\langle\sigma v\rangle$ and use this to calculate the 95\% upper confidence limit compared to a null DM contribution (i.e., we take $\chi^2_\text{DM} = \chi^2_\text{DM=0} + 3.84$ which corresponds to the 95\% upper limit on a $\chi^2$ distribution with one degree of freedom).

\section{Results}
\label{sec:results}

In Table~\ref{tab: parameters} we show the best-fitting parameters of our astrophysical background model to the combined AMS-02 positron, proton and He data (Step 2). In general, our best fit parameters for the diffusion coefficients, proton injection spectral indices, the z-dependence of the convection velocity, Alfv\'enic velocity and solar modulation parameters are consistent with previous analyses~\cite{Trotta:2010mx, Johannesson:2016rlh, Korsmeier:2021brc, Cholis:2017qlb}. We note that our pulsar formation rate is about an order of magnitude smaller as e.g. in Ref.~\cite{Hooper:2008kg}. However, this value depends strongly on the assumptions made about the pulsars (e.g., total spin-down luminosity and initial period), as well as the spectrum, which controls the fraction of the pulsar spin-down power that is injected into the critical 10--1000~GeV range where the pulsar contribution is significant. Several of these parameters (e.g., the total spin-down luminosity) have order of magnitude uncertainties. Another notable result in our fit is that the break in the proton injection spectrum  is positioned at a few GeV, which is very close to the lower limit of our model. This means that a break in the proton injection spectrum is not heavily favored in our model.

In Table~\ref{tab: chi^2 summary}, we show the resulting fits for both of our models to the combined AMS-02 dataset, as well as the individual fits for each cosmic-ray species.  For the default model with a standard halo height of 5.6~kpc, the total $\chi^2$ is 88.60 over 141 degrees of freedom. The $\chi^2$ per degree of freedom for positrons, protons and He are 0.88, 0.43, and 0.57, respectively. The $\chi^2$/d.o.f. is less than 1 for each model and species, indicating that our model fits the data very well. We note that we treat the systematic uncertainties of nearby energy bins as uncorrelated, even though they are likely to be correlated, which results in an overestimation of the uncertainties and thus a $\chi^2$/d.o.f. below 1. The total $\chi^2$ is better in the default model than the small-height model by $\sim3\sigma$, which means that 3~kpc approximately represents the minimum halo height that is consistent with AMS-02 data. 

\begin{figure}[tbp]
\includegraphics[width=0.45\textwidth]{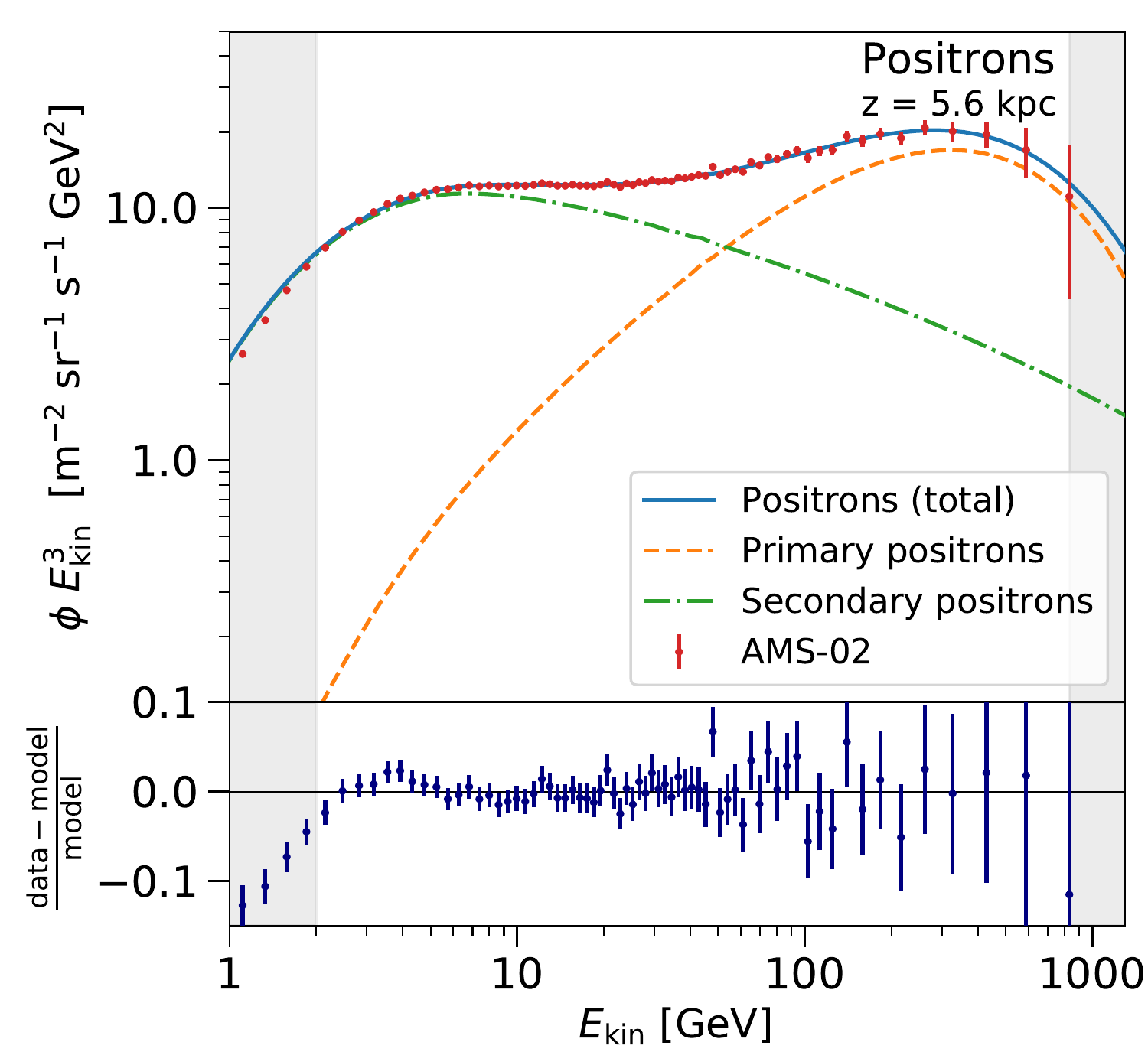}
\hspace{1.5cm}
\includegraphics[width=0.45\textwidth]{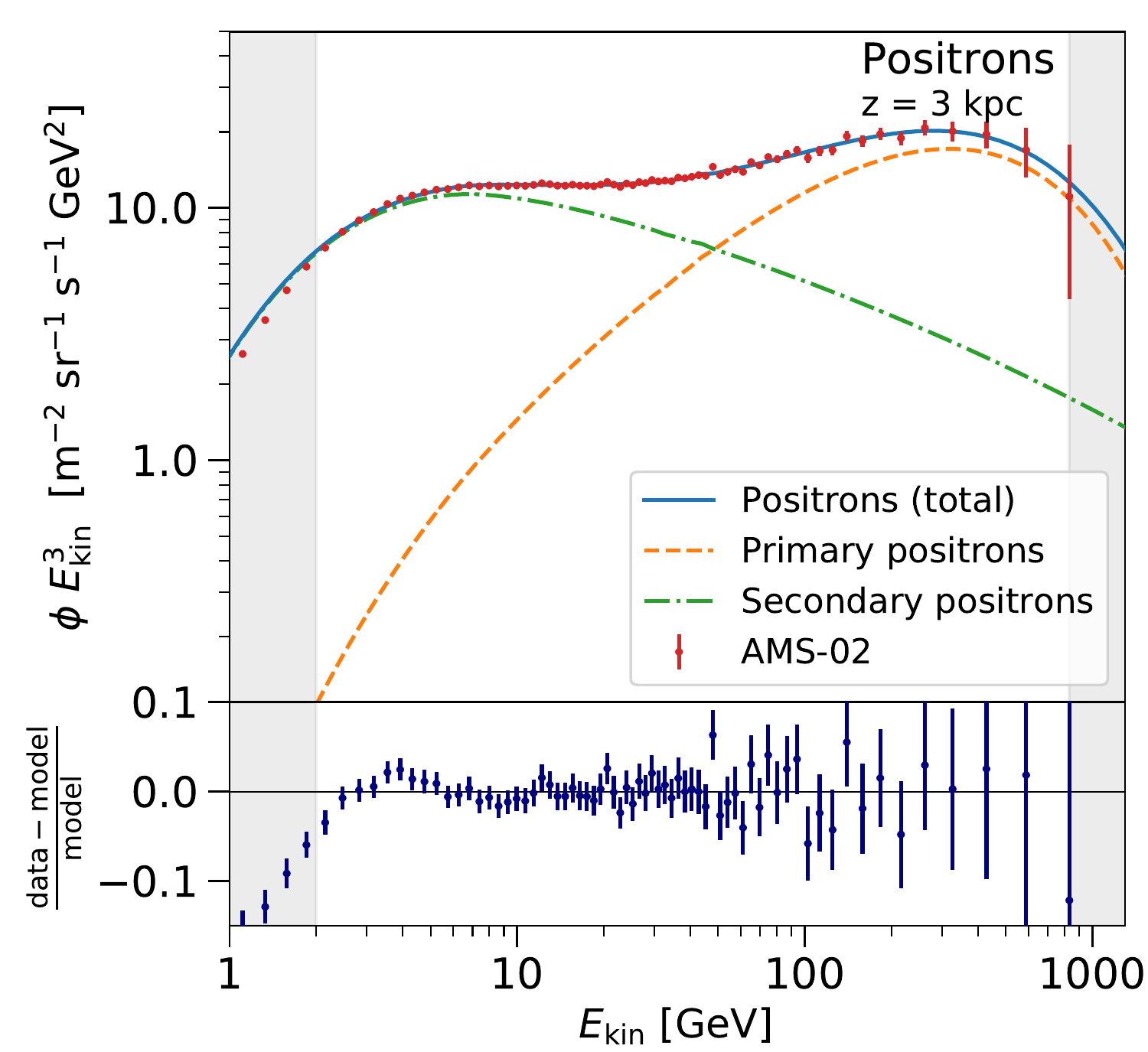}
\includegraphics[width=0.45\textwidth]{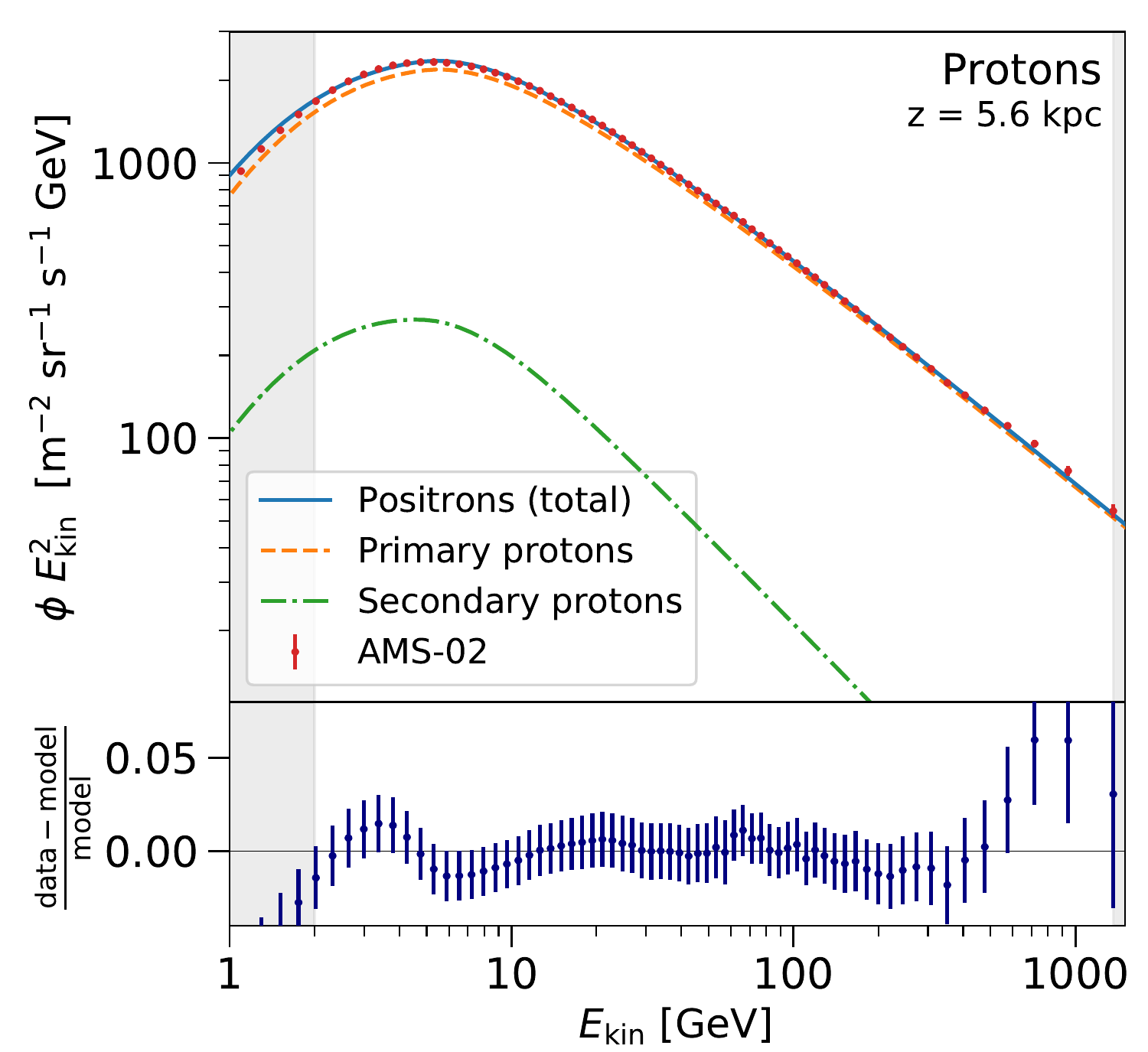}
\hspace{1.5cm}
\includegraphics[width=0.45\textwidth]{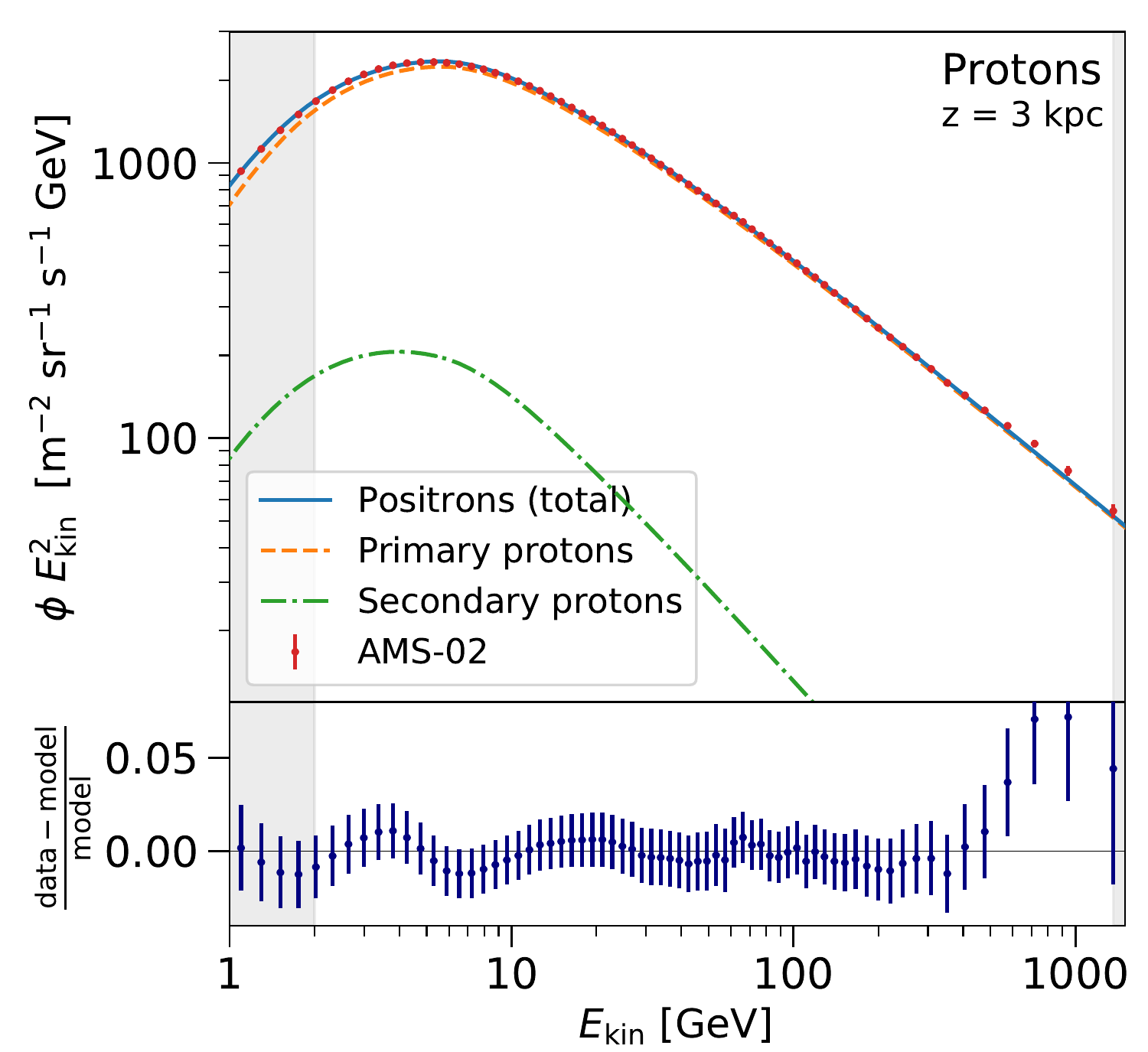}
\includegraphics[width=0.45\textwidth]{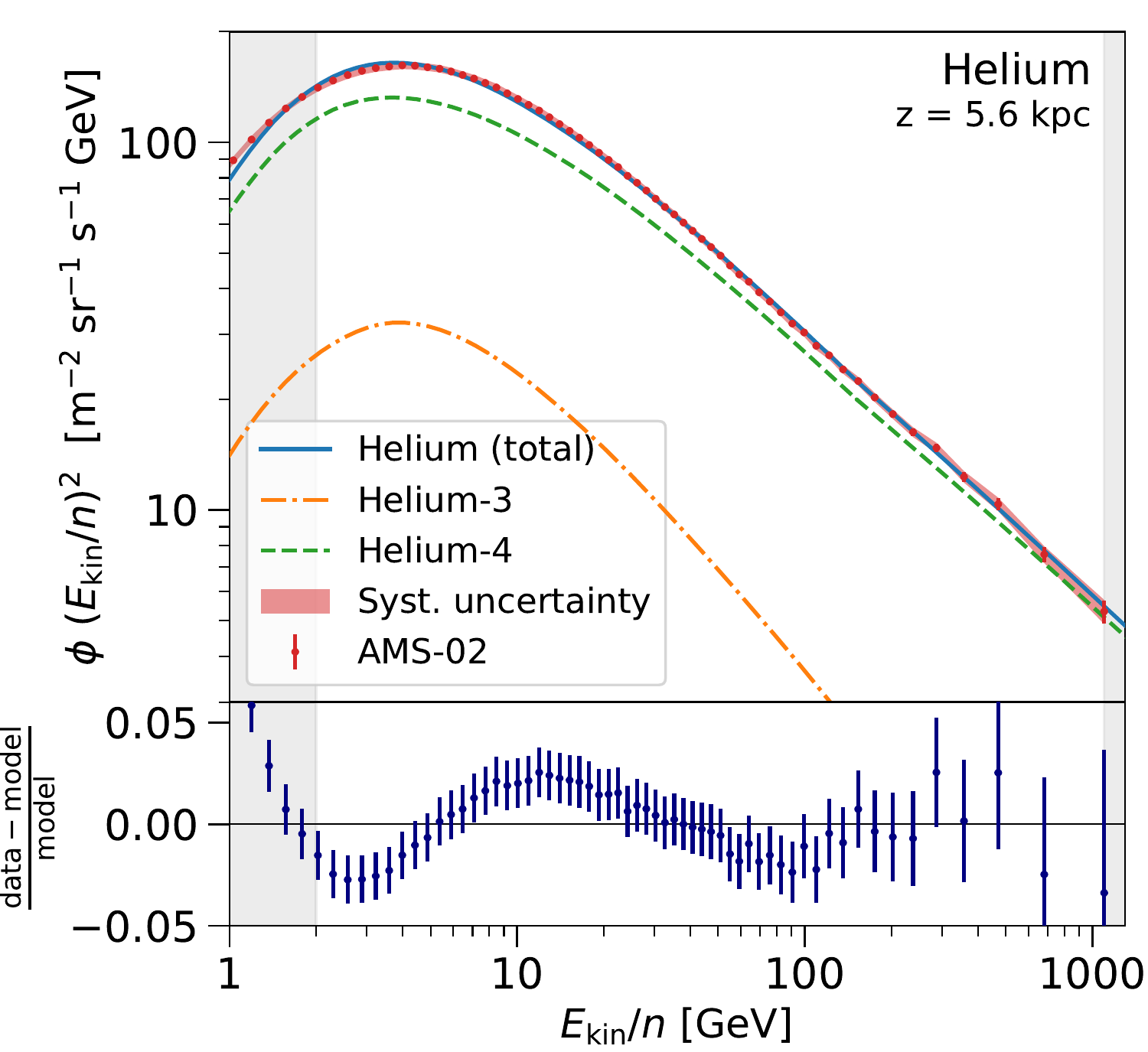}
\hspace{1.5cm}
\includegraphics[width=0.45\textwidth]{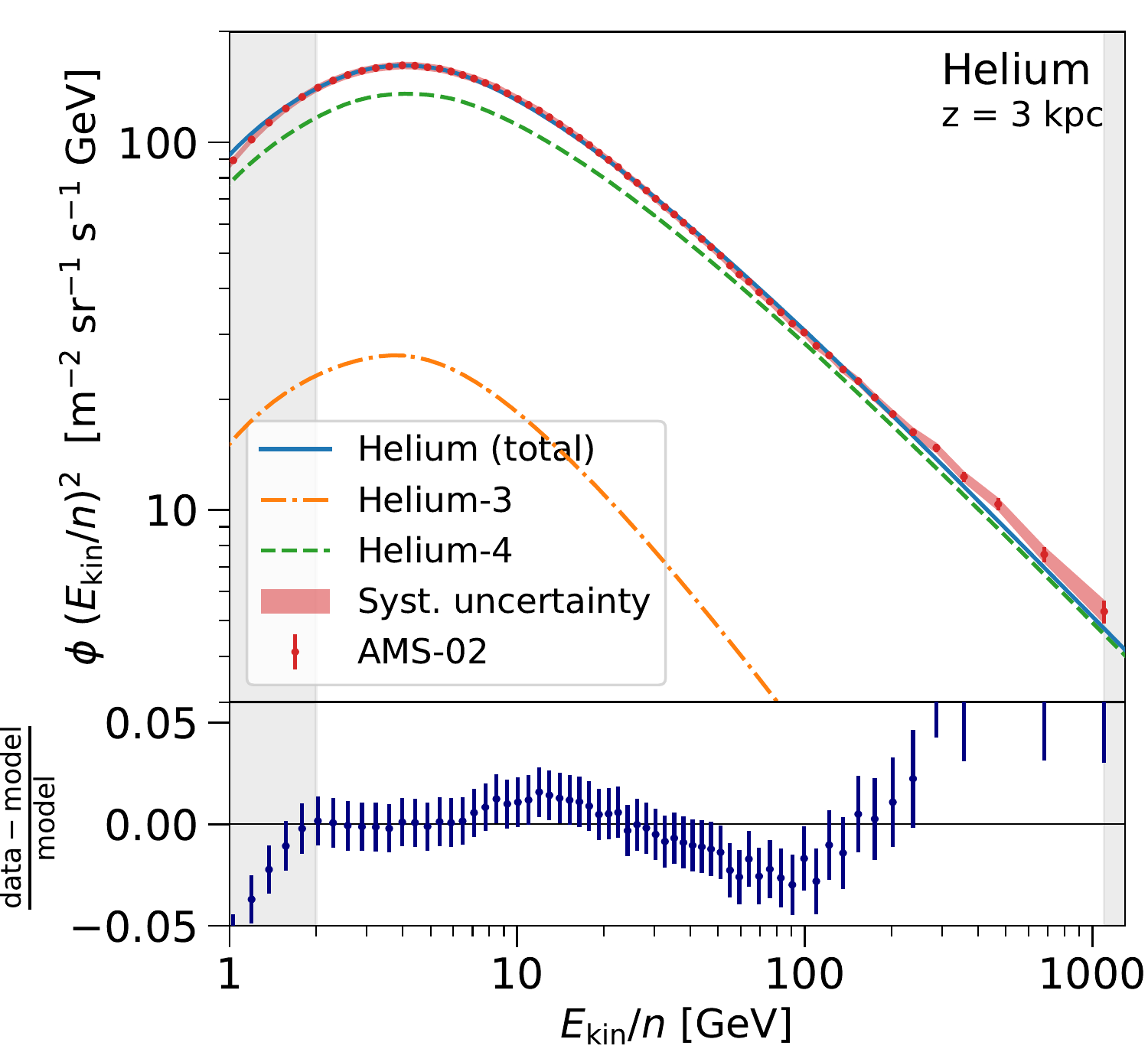}
\caption{The astrophysical model for positrons (top), protons (middle), and He (bottom), plotted in solid blue lines, together with the data from AMS-02 for positrons \cite{PhysRevLett.122.041102}, protons \cite{PhysRevLett.114.171103, PhysRevLett.121.051101} and He \cite{PhysRevLett.115.211101}, displayed in red. The left panels show the default model (5.6~kpc halo height) and the right panels the small-height model (3~kpc halo height). For positrons and protons, primary components are displayed in orange (dashed), secondary components in green (dashed-dotted). $^3$He and $^4$He are represented by orange (dashed-dotted) and green (dashed), respectively. Note also that the He fit includes an additional systematic uncertainty band of 2\%. Gray shaded regions are excluded from the fit. The fractional residuals of the fits are given at the bottom of each panel, showing that we fit all data to a few percent accuracy.}
\label{fig: astrophysical model iminuit}
\end{figure}

In Figure~\ref{fig: astrophysical model iminuit} we show our best-fit model for positrons, protons and He, comparing our results to the AMS-02 data for the best-fit model parameters given in Table~\ref{tab: parameters}. In the bottom of each plot in Figure~\ref{fig: astrophysical model iminuit}, we give the residuals ((data-model)/model) of each fit, showing that our model fits the data to within a few percent, providing a very good fit to the cosmic-ray data that are relevant to the positron flux. We observe some systematic uncertainties for low-energy positrons (below 6~GeV) and protons and He at higher energies (above 600~GeV). This may correspond to additional secondary proton reacceleration in supernova remnants that is not taken into account in Galprop~\cite{Cholis:2017qlb}. Note that the gray shaded regions below 2~GeV and above about 800~GeV, 1050~GeV and 1000~GeV for positrons, protons and He, respectively, are not included in the fit. This means that our model has larger uncertainties at energies near these limits. 

We point out that our analysis does not take into account the correlation of the systematic uncertainties between nearby energy bins in the AMS-02 data. As can be observed in the residual plots in Figure~\ref{fig: astrophysical model iminuit}, especially for protons and He, the best-fit residual values are highly correlated between nearby energy bins, pointing towards a strong correlation in the systematic uncertainties. We are not able to take these correlations into account because the covariance matrices are not provided by AMS-02, which leads to a miss-estimation of the uncertainties. We additionally note that our calculation of the $\chi^2$/d.o.f.~for He is artificially decreased by the additional 2\% systematic error added into the He uncertainties. However, we stress that the best-fit value of our parameters, as well as the value of $\Delta\chi^2$ between our models, is preserved in light of these systematic uncertainties.

\begin{table}[tbp]
\centering
\begin{tabular}{|c|c|c|c|} 
\hline
\rule{0pt}{0.3cm}
~Halo height~ & Type & $\chi^2$ & ~$\chi^2$/d.o.f. (d.o.f.)~ \\
\hline\hline
\rule{0pt}{0.3cm}
\multirow{4}{*}{5.6 kpc} & total & ~88.60~ & ~0.63 (141)~\\
 & ~positrons~ & 42.88 & 0.88 (49)\\
 & protons     & 21.08 & 0.43 (49)\\
 & H           & 24.64 & 0.57 (43)\\
\hline
\rule{0pt}{0.3cm}
\multirow{4}{*}{3 kpc} & total & 97.08 & 0.69 (141) \\
 & positrons & 47.70 & 0.97 (49) \\
 & protons   & 18.97 & 0.39 (49) \\
 & H         & 30.41 & 0.71 (43) \\
\hline
\end{tabular}
\caption{The $\chi^2$-values and $\chi^2$ per degree of freedom of the iMinuit fit for the total flux and the individual fluxes for positrons, protons and He. Values for the default model are given in the top rows and for the small-height model in the bottom rows.}
\label{tab: chi^2 summary}
\end{table}

\begin{figure}[tbp]
\centering
\includegraphics[width=0.48\textwidth]{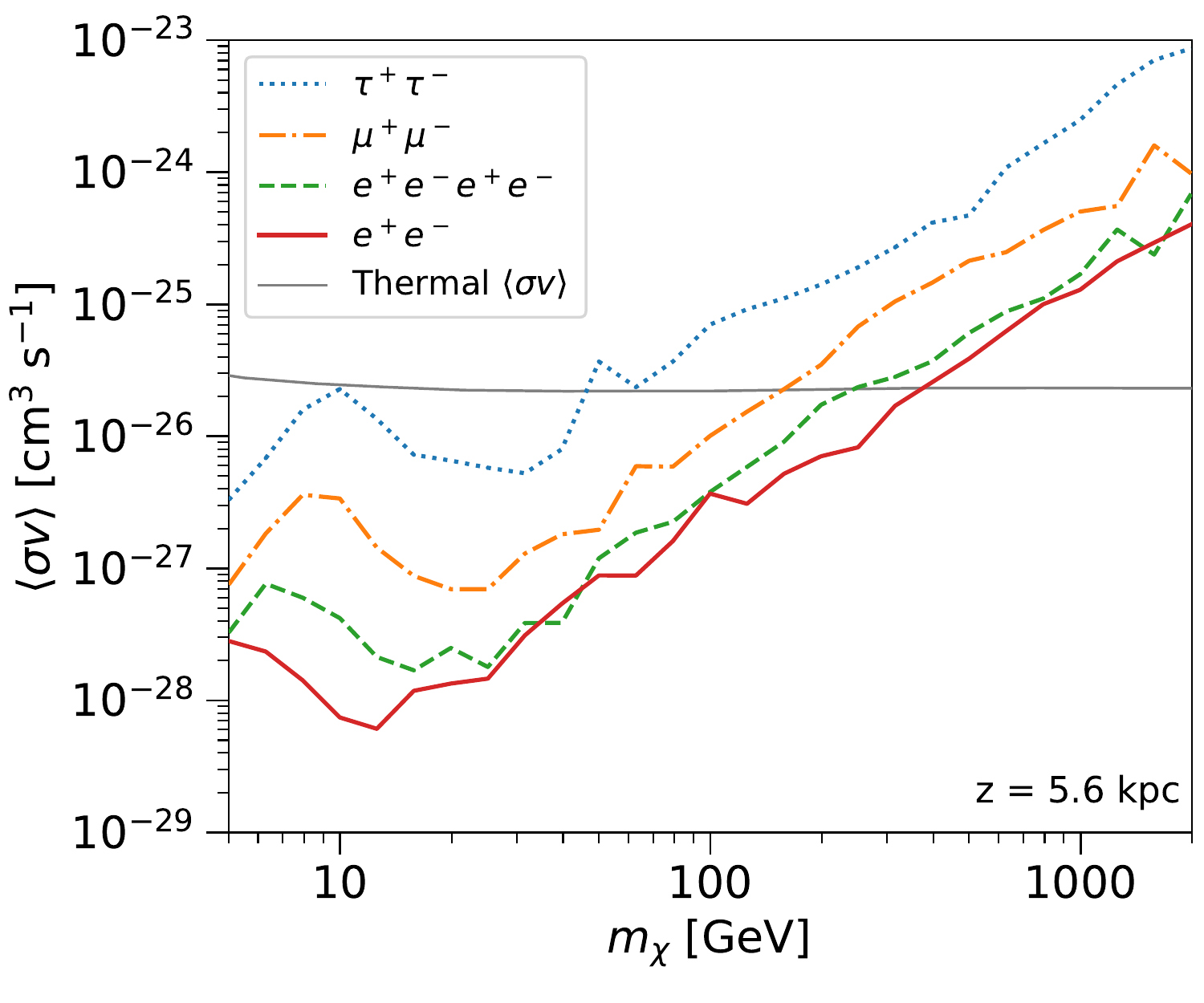}
\hspace{0.3cm}
\includegraphics[width=0.48\textwidth]{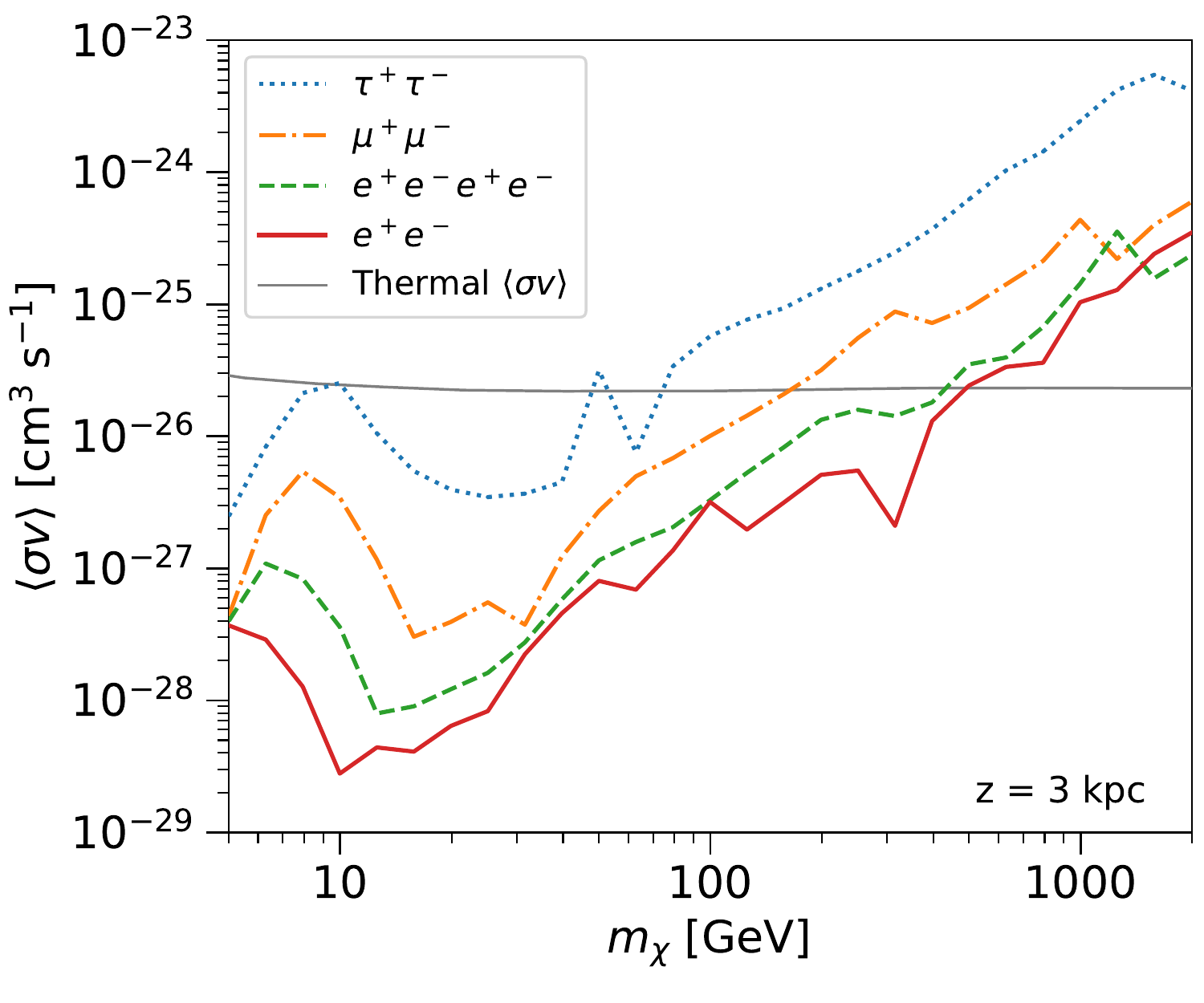}
\caption{DM limits for different leptophilic final states for a DM mass range from 5 to 2000~GeV, for a halo height of 5.6~kpc (left) in the default model and 3~kpc (right) in the small-height model, for decays into $\tau^+\tau^-$ (blue, dotted), $\mu^+\mu^-$ (orange, dashed-dotted), $e^+e^-e^+e^-$ (green, dashed) and $e^+e^-$ (red, solid). The thermal cross section (gray, solid) is taken from \cite{Steigman:2012nb}. Our limits rule out a thermal annihilation cross section for DM masses below between 60 to 300~GeV in the default model and between 60 and 500~GeV in the small-height model, depending on the final state.}
\label{fig: DM limits}
\end{figure}

In Appendix \ref{app: Positron Flux with DM Contribution}, we show an example of the positron flux including a DM contribution for annihilations into $e^+e^-$ at 5~GeV where the excess is largest.

In Figure~\ref{fig: DM limits}, the 95\% confidence upper limits on the DM annihilation cross-sections are presented for the various final decay states. The left panel shows the results of the default model and the right panel the small-height model. For reference, we also include the thermal cross section from \cite{Steigman:2012nb} (for a more recent analysis, see also \cite{Finkel:2020lgf}). We derive the strongest constraints for DM annihilation directly into $e^+e^-$ pairs with an annihilation cross section of $6\cdot10^{-29}$~cm$^3$/s for 12.6~GeV in the default model, and an annihilation cross section of $2.6\cdot10^{-29}$~cm$^3$/s for 10.0~GeV in the small-height model, both of which lie below the thermal annihilation cross section by about 3 orders of magnitude.
 
Our limits show a small bump at low energies for all final states. Specifically, in the default model for the $\tau^+\tau^-$ final state, the bump peaks at 10~GeV and corresponds to a preference for DM at a local significance of 2.8$\sigma$, for $\mu^+\mu^-$ the peak is at 8~GeV with a local significance of 2.9$\sigma$, for $e^+e^-e^+e^-$ the peak is at 6.3~GeV with a local significance of 2.8$\sigma$, and for $e^+e^-$ the peak is at 5~GeV with a local significance of 3.1$\sigma$. In the small-height model, for $\tau^+\tau^-$ the excess peaks at 13~GeV and 2.7$\sigma$, for $\mu^+\mu^-$ at 8~GeV and 3.9$\sigma$, for $e^+e^-e^+e^-$ at 6~GeV and 3.6$\sigma$ and for $e^+e^-$ at 5~GeV and 3.6$\sigma$. These all correspond to the same spectral feature in the low energy positron fit of our model, which can be seen in Figure~\ref{fig: astrophysical model iminuit} as a positive discrepancy (about~2\%) between our model and the AMS-02 data below about 5~GeV. We note that this region of our model is prone to larger uncertainties, as these energies are close to the lower limit of our model (2~GeV). Additionally, lower energies are highly affected by uncertainties in solar modulation. Note that the significance of these excesses are local, and we have not implemented a proper trials correction here (as e.g. in \cite{Conrad:2014nna}) which would reduce the local significance of these excesses. Following the method in \cite{Conrad:2014nna} requires the covariance matrices of the experimental data to simulate pseudo-experiments, but these covariance matrices are not supplied by the AMS-02 experiment. Nonetheless we estimate the correction by counting the number of upcrossing in the residuals of the positron flux, following the method in \cite{Conrad:2014nna}. We count 17 upcrossing which reduces the significances from about 3.0$\sigma$ to about 2.0$\sigma$, and from 3.64$\sigma$ to 2.83$\sigma$. However, we note that the significance of this result is likely much lower, due to the fact that this excess appears very close to the boundary of our modeling, where our models of solar modulation and cosmic-ray propagation are likely beginning to break down.

We point out that recent studies \cite{Genolini:2017dfb, Korsmeier:2021brc} of the cosmic-ray data suggest an additional break in the diffusion spectrum or the particle injection spectra at a few hundred GV that we do not consider here but could potentially strengthen our limits.

\section{Discussion and Conclusions}
\label{sec:discussion}
In this paper, we calculated constraints on the DM annihilation cross section for DM masses between 5 and 2000~GeV for leptonic final states using the local positron flux. We exclude the thermal cross-section for annihilation into $\tau^+\tau^-$ below 60~GeV, for $\mu^+\mu^-$ below 160~GeV, for $e^+e^-e^+e^-$ below 240~GeV and for $e^+e^-$ below 380~GeV for our default model with a halo height of 5.6~kpc.

We repeated our analysis with a conservative halo height of 3~kpc in the small-height model, for which we obtain similar constraints: Our limits are below the thermal cross section for annihilation into $\tau^+\tau^-$ below 50~GeV, for $\mu^+\mu^-$ below 170~GeV, for $e^+e^-e^+e^-$ below 430~GeV and for $e^+e^-$ below 500~GeV. We note that our constraints at higher energies are stronger in the small-height model than the default model. This is because our two models have a different diffusion coefficient at high energies, meaning that the diffusion distance is slightly longer in the small-height model which results in the stronger constraints at high masses ($\gtrsim$~300~GeV). We conclude from our results that a change in halo height does not significantly affect the local positron flux, contrary to limits on the local antiproton flux that depends on the halo height~\cite{Korsmeier:2021brc}. This is due to the fact that the average positron travels a shorter distance than antiprotons at similar energy (because their energy losses are faster) -- making the edges of the diffusion zone less important for leptonic DM limits.

Our analysis improves on the constraints by Bergstr\"om et al. by about a factor of 2 for both models. This is despite the fact that our model gives significantly more freedom to the astrophysical background fit -- and thus includes more conservative and robust limits. Note that Bergstr\"om et al. use a halo height of 4~kpc. We are also able to extend the limits up to DM masses of 2000~GeV as more cosmic-ray data is available at higher energies.

Our limits also set strong constraints on DM models with only subdominant annihilations into the four leptonic final states considered in this work. At about 30~GeV, our limits fall below the thermal cross section by about two orders of magnitude for annihilation into $e^+e^-$. This means that our limits even rule out subdominant contributions from DM annihilations (i.e.,~$\sim$1\% annihilation into $e^+e^-$ pairs at 30~GeV), and similarly $\sim$10\% for annihilations into $\mu^+\mu^-$. Notably, DM models with significant annihilation fractions into $\mu^+\mu^-$ final states are among the most difficult thermal DM models to constrain with current experiments~\cite{Leane:2018kjk}. Compared to the results of Ref.~\cite{Leane:2018kjk}, which only ruled out thermal annihilation to $\mu^+\mu^-$ up to a DM mass of 30~GeV, we rule out annihilation to $\mu^+\mu^-$ up to masses of $\sim$150~GeV. We note that a significant fraction of this difference pertains to less conservative model assumptions utilized in this work, compared to that of Ref.~\cite{Leane:2018kjk}. For example, they use a local DM density of 0.25~GeV/cm$^3$ and local magnetic field strength of 8.9~$\mu$G, both of which serve to significantly decrease the local positron flux from DM at a given annihilation cross-section. On the other hand, our limits benefit from the use of a physical model for the astrophysical contributions from pulsars and secondary production, rather than producing an empirical fit to the data. Our limits also profit from the better data by AMS-02, which dominates the improvement of our limits at higher energies. Thus our results can also improve generic constraints on DM annihilation to mixed final states.

Our limits are highly complementary with $\gamma$-ray constraints on DM annihilation. In particular, direct annihilations to $e^+e^-$, $\mu^+\mu^-$ and light mediator final states are difficult to probe with $\gamma$-ray instruments, because their decays produce few $\gamma$-rays at tree-level. On the other hand, $\gamma$-ray studies are highly sensitive to DM annihilation into hadronic final states, which are not tested here. Our limits, in conjunction with current limits on DM annihilation to hadronic states (e.g., in dwarf spheroidal galaxies) place strong limits on DM annihilation regardless of final state.

\vspace{-0.2cm}
\section*{Acknowledgements}
\vspace{-0.35cm}
We thank John Beacom, Ilias Cholis, Carmelo Evoli, Andrei Kounine and Rebecca Leane for comments that improved the quality of this paper. We would like to especially thank Michael Korsmeier for numerous comments and discussions regarding the production of our astrophysical models. TL is supported by the Swedish Research Council under contract 2019-05135,  the Swedish  National  Space  Agency  under contract 117/19 and the European Research Council under grant 742104. This project used computing resources from the Swedish National Infrastructure for Computing (SNIC) under project Nos. 2020/5-463 and 2021-1-24 partially funded by the Swedish Research Council through grant no. 2018-05973.

\bibliography{main}{}
\bibliographystyle{JHEP}

\appendix
\newpage

\section{Astrophysical Model with Multinest}
\label{sec:multinest}
In this section, we present the results from the first step of creating our astrophysical background models, in which we fit our Galprop model to the AMS-02 data for positrons and protons (see Table~\ref{tab: AMSdata} for the specific AMS-02 data sets) with \texttt{Multinest} \cite{pymultinest}, i.e., we do not fit the He spectrum in this step. We choose to include this step because Multinest is very reliable at finding the global minimum of the fit, whereas iMinuit is more likely to fall into local minima in large multi-dimensional parameter spaces. The free parameters can be found in Tables~\ref{tab: multinest parameters} and~\ref{tab: multinest parameters_3kpc} for the default and small-height model, respectively, and include the following parameters: the diffusion spectrum, the proton injection spectrum (with a spectral break and a spectral index below and above the break), the change in convection velocity, the Alfv\'{e}n velocity, the Galactic pulsar formation rate and spectrum (with a spectral index and energy cutoff), the solar modulation parameters and an overall normalization of the primary particle abundance of $1.06\cdot10^6$ for protons in Galprop.

We present our results in Tables~\ref{tab: multinest parameters} and~\ref{tab: multinest parameters_3kpc} for the default and small-height model, respectively, where we give the range in which we allow the parameters to float, the type of prior (i.e.,~ linear or logarithmic), the best-fit value, the posterior value and the posterior uncertainty. In particular, we scan the priors for $D_0$, $D_\text{break}$, the proton injection spectral break and $E^\text{psr}_\text{cut}$ in logarithmic space, while we use a flat distribution for the other parameters. We obtain results that are relatively consistent between the default and small-height model.

\begin{figure}[tbp]
\centering
\includegraphics[width=.48\textwidth]{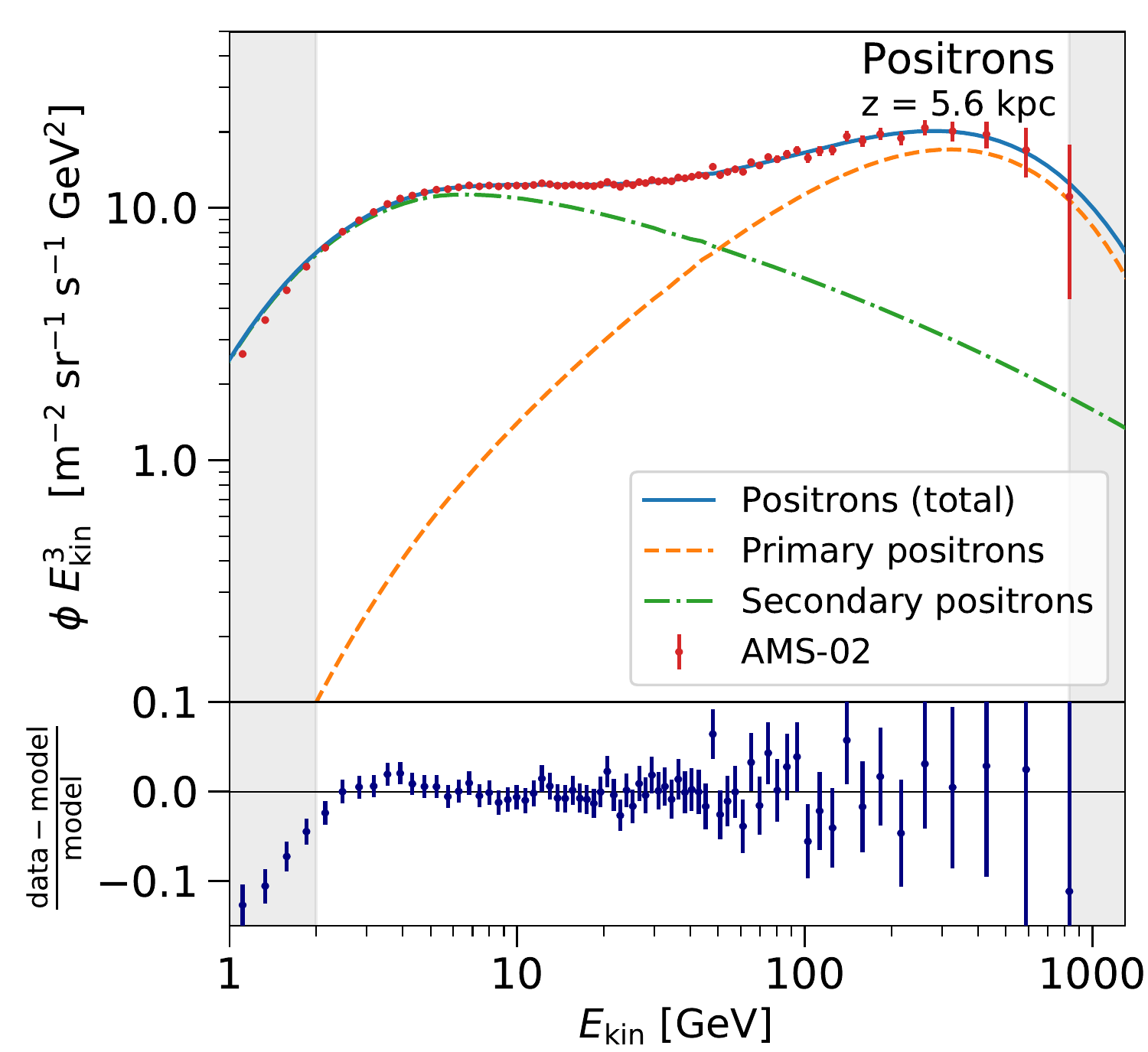}
\hspace{0.3cm}
\includegraphics[width=.48\textwidth]{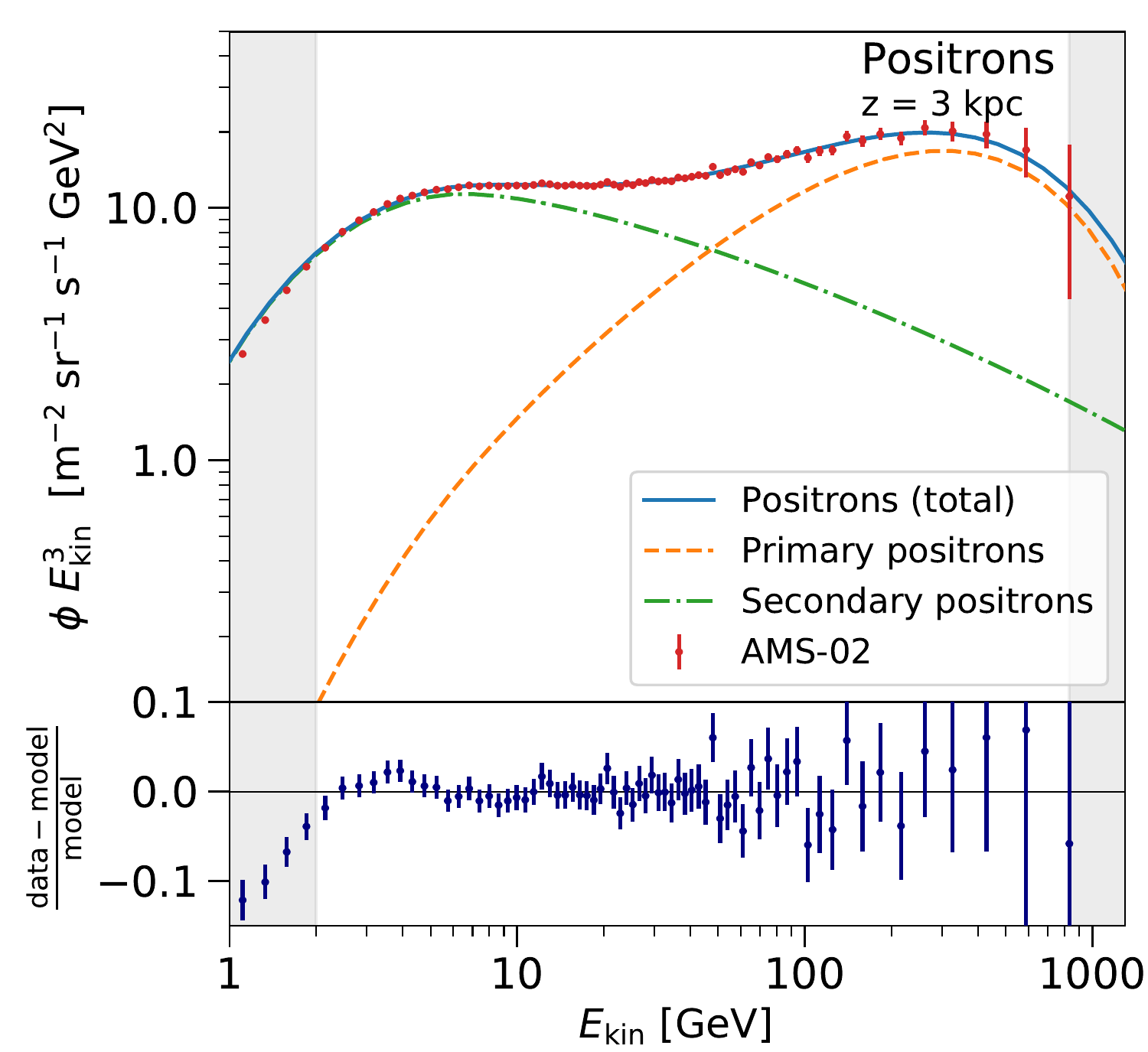}
\includegraphics[width=.48\textwidth]{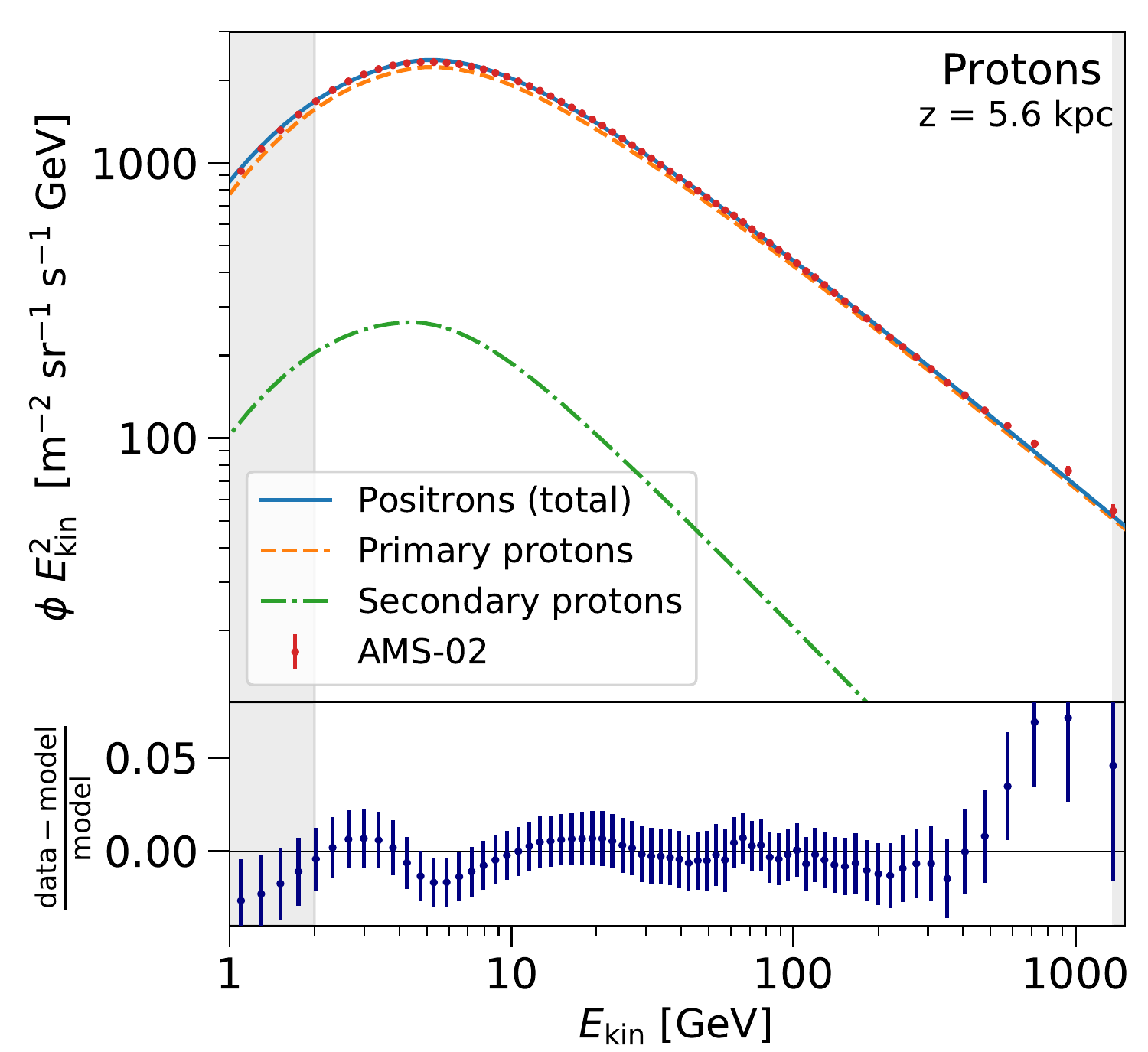}
\hspace{0.3cm}
\includegraphics[width=.48\textwidth]{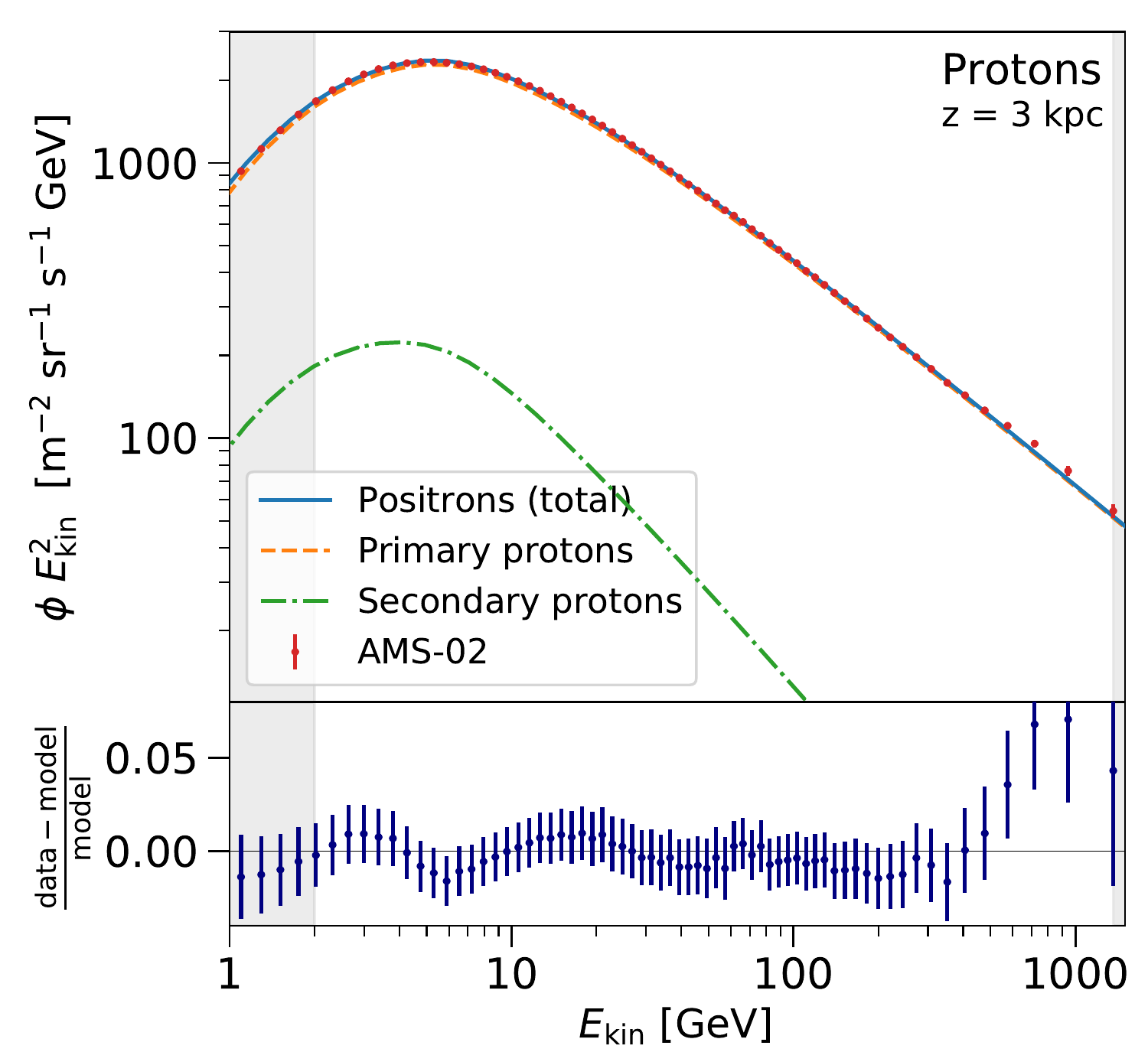}
\caption{Astrophysical model obtained with Multinest for the local flux of positrons (top panels) and protons (bottom panels). Gray shaded regions are excluded when fitting the model to the data. The left panels show the default model ($z=5.6$~kpc), and the right panels show the small-height model ($z=3$~kpc). The AMS-02 data is given in red for positrons \cite{PhysRevLett.122.041102} and protons \cite{PhysRevLett.114.171103, PhysRevLett.121.051101}. The total flux of our model is presented in blue (solid), primary components in orange (dashed) and secondary components in green (dashed-dotted). The fractional residuals between the data and our model are given in the bottom of each panel, showing that our models for positrons and protons fit the data to a few percent accuracy for both halo heights.}
\label{fig:multinest}
\end{figure}

Figure~\ref{fig:multinest} shows our fit for positrons and protons for our default model and the small-height model, together with the AMS-02 data for the local fluxes. The fractional residuals between the data and our model are given in the bottom of each panel, showing that our models fit the AMS-02 data to within a few percent for the data that is relevant to the positron flux.

A summary of the $\chi^2$-values is given in Table~\ref{tab: chi^2 summary multinest} for the default model and small-height model. For the default model, the total $\chi^2$ is 61.5 over 101 degrees of freedom, with $\chi^2_{p}$/d.o.f.~$=0.45$ and $\chi^2_{e^+}$/d.o.f.~$=0.76$. For the small-height model, the total $\chi^2$ is 63.2 over 101 degrees of freedom, with $\chi^2_{p}$/d.o.f.~$=0.45$ and $\chi^2_{e^+}$/d.o.f.~$=0.79$.

The best-fit values (rather than the posterior means) obtained in this step serve as the initial values for the iMinuit fit in the second step, allowing us to start the fitting procedure close to the global minimum. Comparing our Multinest fit with the second iMinuit fit (see Table~\ref{tab: parameters}), we remark that the positron and proton fit only worsen slightly ($\Delta\chi^2$/d.o.f.~$\approx 0.04$) when introducing the fit to the He data in the default model. In the small-height model, $\Delta\chi^2$/d.o.f.~$\approx 0.18$ for positrons, while the proton fit improves by $\Delta\chi^2$/d.o.f.~$\approx 0.06$.

\begin{table}[tbp]
\footnotesize
\begin{tabular}{|c|c|c|c|c|c|} 
\hline
Parameter  & Range & Prior & Best fit & Posterior & Post. unc.\\  
\hline\hline
\rule{0pt}{0.35cm}
Diffusion coefficient, $D_0$ [cm$^2$/s]                  & $10^{28} - 10^{30}$             & log    & $1.563\cdot10^{28}$  & $1.732\cdot10^{28}$ & $1.026\cdot10^{28}$   \\
Break in diff. spectrum, $D_\text{break}$ [MV]     & $10^3 - 1.4 \cdot 10^4$         & log    & $5.479\cdot 10^3$    & $5.758\cdot 10^3$  & $0.283\cdot 10^3$  \\
Spectral index below break, $\delta_1$                   & $0.01 - 0.61$                   & linear & 0.053                & 0.047 & 0.017  \\
Spectral index above break, $\delta_2$                   & $0.01 - 0.61$                   & linear & 0.364                & 0.383 & 0.027  \\
Change in $v_\text{conv}$, $dv/dz$ [km/(s kpc)]                  & $0 - 10$                        & linear & 7.282                & 6.7 & 2.0  \\
Alfv\'en velocity, $v_\text{Alfv\'en}$ [km/s]            & $1 - 30$                        & linear & 4.553                & 4.2 & 1.0  \\
$p$ injection spectrum break [MV]                     & $5\cdot 10^2 - 1\cdot 10^5$ & log    & $2.064\cdot10^3$     & $1.389\cdot10^3$ & $0.787\cdot10^3$  \\
$p$ spectral index below break, $\gamma_1^p$          & $1.5 - 3.0$                     & linear & 2.196                & 2.12 & 0.39  \\
$p$ spectral index above break, $\gamma_2^p$          & $1.5 - 4.5$                     & linear & 2.535                & 2.510 & 0.023  \\
$p$ spectral index, $\gamma_\text{psr}$               & $0 - 2.5$                       & linear & 1.374                & 1.480 & 0.051  \\
Pulsar cutoff energy, $E_\text{cut}^\text{psr}$ [GeV]    & $5\cdot 10^2 - 2\cdot 10^3$     & log    & 554.341              & 714 & 118  \\
Pulsar form. rate, $\dot{N}_{100}$ [psr/century]     & $0.148 - 44.40$                 & linear & 0.297                & 0.497 & 0.132  \\
Solar mod. parameter, $\phi_0$ [GV]                & $0.21 - 0.43$                   & linear & 0.386                & 0.366 & 0.012  \\
Solar mod. parameter, $\phi_1$ [GV]                & $1.15 - 1.95$                   & linear & 1.501                & 1.64 & 0.23  \\
Normalization (positrons, protons)                       & $0.5 - 1.5$                     & linear & 0.818                & 0.8206 & 0.0038  \\
\hline
\end{tabular}
\caption{List of free parameters of our Multinest fit for the default model ($z = 5.6$ kpc). We give the range in which we let the parameters float, the type of prior, the best-fit value, the posterior and the posterior uncertainty. This model provides a good fit to the data with a total $\chi^2$ of 61.54 over 101 degrees of freedom.}
\label{tab: multinest parameters}
\end{table}

\newpage 

\begin{table}[tbp]
\footnotesize
\begin{tabular}{|c|c|c|c|c|c|} 
\hline
Parameter  & Range & Prior & Best fit & Posterior & Post. unc.\\  
\hline\hline
\rule{0pt}{0.35cm}
Diffusion coefficient, $D_0$ [cm$^2$/s]                  & $10^{28} - 10^{30}$             & log    & $1.822\cdot10^{28}$  & $1.994\cdot10^{28}$ & $1.316\cdot10^{28}$  \\
Break in diff. spectrum, $D_\text{break}$ [MV]     & $10^3 - 1.4 \cdot 10^4$         & log    & $5.938\cdot 10^3$    & $6.026\cdot 10^3$  & $0.371\cdot 10^3$   \\
Spectral index below break, $\delta_1$                   & $0.01 - 0.61$                   & linear & 0.147                & 0.123 & 0.038  \\
Spectral index above break, $\delta_2$                   & $0.01 - 0.61$                   & linear & 0.430                & 0.460 & 0.037  \\
Change in $v_\text{conv}$, $dv/dz$ [km/(s kpc)]                  & $0 - 10$                        & linear & 6.724                & 2.4 & 2.0  \\
Alfv\'en velocity, $v_\text{Alfv\'en}$ [km/s]            & $1 - 30$                        & linear & 9.148                & 6.8 & 1.4  \\
$p$ injection spectrum break [MV]                     & $5\cdot 10^2 - 1\cdot 10^5$ & log    & $1.649\cdot10^3$     & $1.335\cdot10^3$  & $0.841\cdot10^3$   \\
$p$ spectral index below break, $\gamma_1^p$          & $1.5 - 3.0$                     & linear & 1.546                & 2.19 & 0.43  \\
$p$ spectral index above break, $\gamma_2^p$          & $1.5 - 4.5$                     & linear & 2.429                & 2.389 & 0.034  \\
Pulsar spectral index, $\gamma_\text{psr}$               & $0 - 2.5$                       & linear & 1.398                & 1.487 & 0.057  \\
Pulsar cutoff energy, $E_\text{cut}^\text{psr}$ [GeV]    & $5\cdot 10^2 - 2\cdot 10^3$     & log    & 506.825              & 640 & 108  \\
Pulsar form. rate, $\dot{N}_{100}$ [psr/century]     & $0.148 - 44.40$                 & linear & 0.351                & 0.479 & 0.171  \\
Solar mod. parameter, $\phi_0$ [GV]                & $0.21 - 0.43$                   & linear & 0.403                & 0.377 & 0.023  \\
Solar mod. parameter, $\phi_1$ [GV]                & $1.15 - 1.95$                   & linear & 1.402                & 1.54 & 0.26  \\
Normalization (positrons, protons)                       & $0.5 - 1.5$                     & linear & 0.824                & 0.8266 & 0.0037  \\
\hline
\end{tabular}
\caption{Same as Table~\ref{tab: multinest parameters}, but for the small-height model ($z=3$~kpc). This model provides a good fit to the data, with a $\chi^2$ of 63.24 over 101 degrees of freedom. The results are relatively consistent with the default model.}
\label{tab: multinest parameters_3kpc}
\end{table}

\begin{table}[tbp]
\centering
\begin{tabular}{|c|c|c|c|} 
\hline
~Halo height~ & Type & $\chi^2$ & ~$\chi^2$/d.o.f. (d.o.f.)~ \\
\hline\hline
\multirow{3}{*}{5.6 kpc} & total & ~61.54~ & ~0.61 (101)~\\
 & ~positrons~ & 39.41 & 0.76 (52)\\
 & protons     & 22.13 & 0.45 (49)\\
 \hline
\multirow{3}{*}{3 kpc} & total & 63.24 & 0.63 (101)\\
 & positrons & 41.22 & 0.79 (52) \\
 & protons   & 22.02 & 0.45 (49) \\
\hline
\end{tabular}
\caption{Summary of the $\chi^2$-values of the Multinest fit for the default model (halo height 5.6~kpc, top rows) and the small-height model (halo height 3~kpc, bottom rows) for the total flux, as well as the individual positron and proton flux. All $\chi^2$ per degree of freedom are below 1, indicating very good agreement of our astrophysical model with the AMS-02 data.}
\label{tab: chi^2 summary multinest}
\end{table}

\clearpage
\section{Positron Flux with Dark Matter Contribution}\label{app: Positron Flux with DM Contribution}
\begin{figure}[h]
\centering
\includegraphics[width=.5\textwidth]{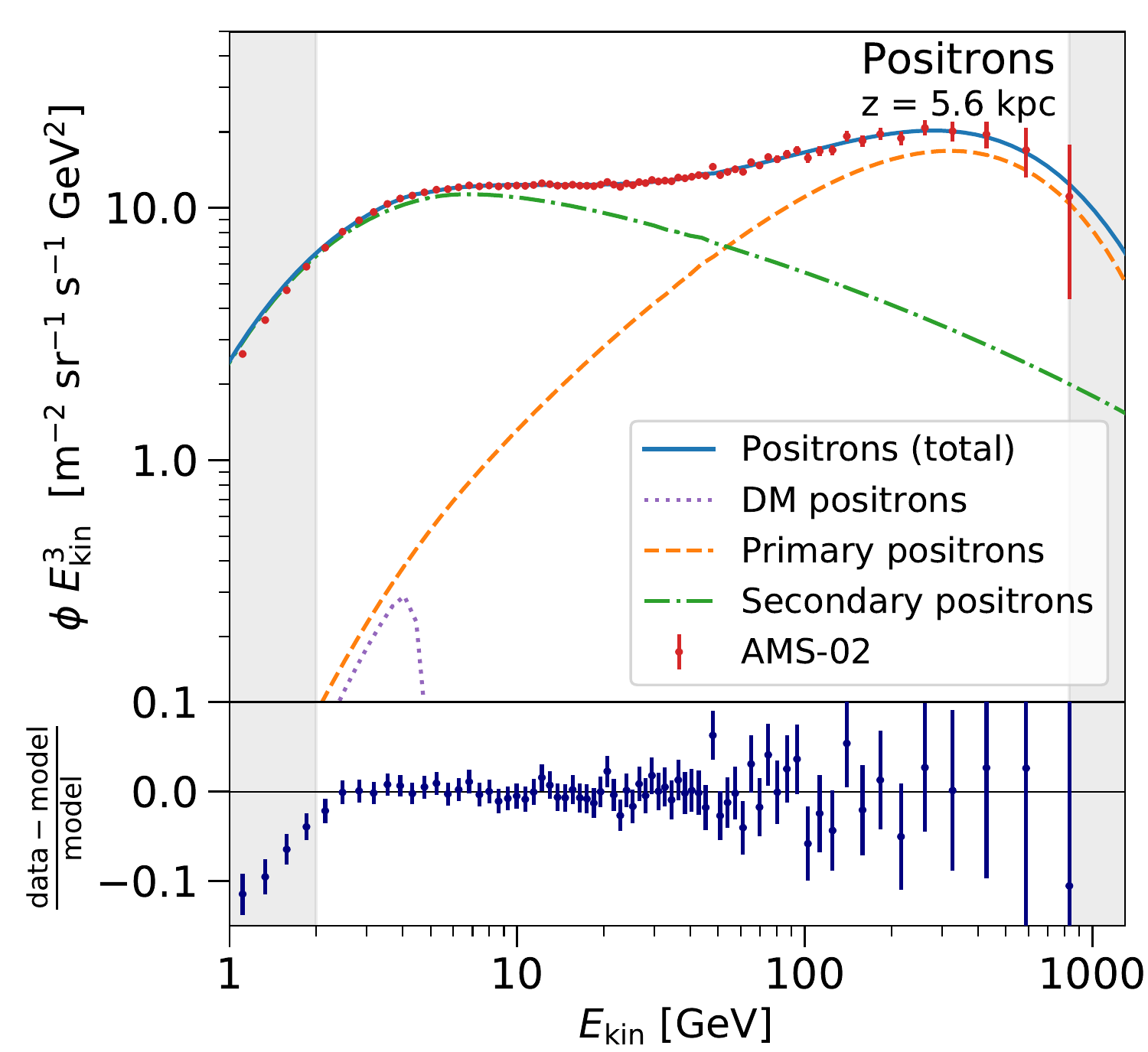}
\caption{Total positron flux (solid, blue) with the components of the background model (primary and secondary positrons) and a contribution from 5-GeV DM annihilating into $e^+e^-$ (purple, dotted) for our default model.}
\label{fig:flux with dm}
\end{figure}

In Figure~\ref{fig:flux with dm} we show an example of the local positron flux as obtained from our simulations with a contribution from 5-GeV DM annihilating into the $e^+e^-$ for our default model.

\end{document}